\documentclass[a4paper,onecolumn,fleqn,usenatbib]{mnras}

\usepackage[T1]{fontenc}
\usepackage{ae,aecompl}
\usepackage{graphicx}	
\usepackage{amsmath}	
\usepackage{amssymb}	

\title[Long term evolution]{Long term evolution of planetary systems with a terrestrial planet and a giant planet}

\author[Nikolaos Georgakarakos, Ian Dobbs-Dixon, Michael J. Way]{
Nikolaos Georgakarakos,$^{1}$\thanks{E-mail: ng53@nyu.edu, georgakarakos@hotmail.com}
Ian Dobbs-Dixon,$^{1}$
Michael J. Way$^{2}$
\\
$^{1}$New York University Abu Dhabi, Saadiyat Island, P.O. Box 129188, Abu Dhabi, UAE\\
$^{2}$ NASA, Goddard Institute for Space Studies, New York, USA}

\date{Accepted XXX. Received YYY; in original form ZZZ}

\pubyear{2015}

\begin{document}
\label{firstpage}
\pagerange{\pageref{firstpage}--\pageref{lastpage}}
\maketitle

\begin{abstract}
We study the long term orbital evolution of a terrestrial planet
under the gravitational perturbations of a giant planet.  In particular, we are
interested in situations where the two planets are in the same plane and are
relatively close.  We examine both possible configurations: the giant planet
orbit being either outside or inside the orbit of the smaller planet.  The
perturbing potential is expanded to high orders and an analytical solution of
the terrestrial planetary orbit is derived.  The analytical estimates are then
compared against results from the numerical integration of the full equations of motion and
we find that the analytical solution works reasonably well. An interesting finding is that the 
new analytical estimates improve greatly the predictions for the timescales of the orbital evolution of the terrestrial planet compared to an octupole order expansion.
Finally, we briefly discuss possible applications of the analytical estimates in astrophysical problems.
\end{abstract}



\section{Introduction}
Three decades ago, the only planets we knew of were the ones in our Solar System.  Today,  over 3400 planets 
have been discovered in 2551 planetary systems, with 581 of those systems having at least two planets (www.exoplanet.eu).   Contrary to our Solar System where the planets, with the exception of Mercury,
have almost circular and coplanar orbits, the exoplanetary systems that have been discovered so far have
planetary eccentricities which range from zero to almost one, as in the case of HD20782b \citep{2009MNRAS.392..641O}.  Other may have mutually inclined orbits as a possible configuration, as the 
discovery of $\upsilon$  Andromedae with the two giant planets having a $30^{\circ}$ mutual inclination \citep{2010ApJ...715.1203M} shows.  For a nice recent review about the architecture of exoplanetary 
systems, one may read \cite{2015ARA&A..53..409W}.

Normally, a good approach to modelling the long term behaviour of a three body system is that of the hierarchical triple, where the system can be viewed as two interacting binaries; two bodies that are close form the inner binary, while the third body is on a wider orbit with respect to the inner binary barycentre.  In such a case, the perturbing potential may be expressed as a power series of the semi-major axis ratio and can be
truncated to whatever order is considered necessary for the problem studied. For stellar systems or planets in
binaries, it is usual to include terms that arise from the $P_2$ and $P_3$ terms when the perturbing Hamiltonian is expanded in terms of Legendre polynomials (the so called quadrupole and octupole terms), e.g. see 
\cite{Ford-et-al-2000}, \cite{Lee-et-al-2003}, \cite{Georgakarakos-2004}, \cite{Naoz-et-al-2013a}.  That approach is normally sufficient when we have at least two stars in the three body system or two planets in highly inclined orbits. This is because the separations need to be large enough so that the system is stable \citep[e.g. see][]{2013NewA...23...41G}.  However, in an almost coplanar system with only one star, the planets may be much closer and still remain stable.  

In a series of papers \citep{Georgakarakos-2002, Georgakarakos-2003, Georgakarakos-2009} we obtained analytical expressions for the eccentricity and the longitude of pericentre of non-resonant coplanar hierarchical triple systems.
Those calculations were done on both short (of the order of the orbital periods) and secular timescales.  Although there was not any explicit restriction on the mass ratios of the systems, that work was intended to focus on stellar systems.  In \cite{Georgakarakos-2006}, we tested the validity of the results of \cite{Georgakarakos-2002, Georgakarakos-2003} for planetary systems, but only as the mass ratios were concerned and only down to Jupiter sized bodies.  In this work, we attempt to extend our previous work by deriving analytical expressions for the motion of a two planet system with all bodies lying in the same plane.  We are particularly interested in the case where one of the planets is much smaller than the other one, as that configuration can have many applications in habitability studies.  Hence, in the present work we are primarily interested in systems with an Earth-like or super Earth planet and a giant planet. The results may be applicable to the motion of other systems as well. For example a system with a star, a giant planet and its exomoon. Finally, we would like to point out that in the present work we will only focus on the long term motion of the system.  

The structure of the paper is as follows: in section 2, we present the formulation and the analytical solution to our problem.  In section 3, we test the new results using numerical integrations of the full equations of motion and in section 4 we summarise our results and discuss future prospects.

\section{Theory}

As we stated briefly in the introduction, the aim of this paper is to derive analytical estimates for the motion of an Earth or Super-Earth mass planet under the gravitational influence of a giant planet, with all bodies lying in the same plane.  We will focus on the long term orbital evolution, since the contribution of short period terms in the orbital evolution of the type of systems we investigate is quite small (with the exception of almost circular giant planet orbits). Since in the secular problem the semi-major axes of the system do not vary  \citep[e.g. see][]{Harrington-1968}, we only need to determine the eccentricities and longitudes of pericentre of our planets in order to have a complete dynamical description of our system.  Assuming that the orbit of the larger planet is not affected by the presence of the smaller one, we only need to work on the orbital evolution of the latter.

Two different setups are investigated.  In the first one, the inner binary of our hierarchical triple will
consist of the star and the smaller planet, while the larger body will be in an orbit exterior to that. That
case will be called {\it inner case} hereinafter. In the second one called {\it outer case}, we consider the
situation where the giant planet orbit is inside the orbit of the smaller planet.  This is also a possible configuration as shown by both recent studies \citep{2006Sci...313.1413R, 2014ApJ...787..172O} and the discovery of systems such as
Kepler-87 \cite[e.g. see][]{2014A&A...561A.103O}.

In what follows, we shall use the Hamiltonian formulation for studying the secular motion of the systems under investigation.
First, we shall derive the secular Hamiltonian and subsequently the equations of motion
which are then solved analytically, providing us with a solution to our problem.  The orbit of the smaller planet is assumed to be initially circular
and that its maximum eccentricity will remain rather low. 
Additionally, in the orbital solution of the outer case, we keep terms of the eighth order 
in giant planet eccentricity, since most hot/warm Jupiters have small to medium eccentricities. All bodies are treated as point-masses and at this stage we only consider gravitational interactions at the Newtonian level.  For all orbital elements, the subscripts 1 and 2 refer to the inner and outer planet respectively.

\subsection{Averaging of the Hamiltonian}

In our previous work \citep{Georgakarakos-2002, Georgakarakos-2003}, the derivation of the secular Hamiltonian was done by means of the Von Zeipel method, which uses canonical transformations to remove the fast variables.
Besides giving a Hamiltonian which is complete in terms of the eccentricities, the Von Zeipel method also produces terms that are of order two in the masses, something that is not achievable by many other methods (e.g. the 'scissors' method).  However, the latter feature of the Von Zeipel method is important for large outer mass systems, which is not the case here.  Hence, 
for the purpose of this paper we proceed with a different way of obtaining the secular Hamiltonian.

\begin{center}
\begin{figure}
\includegraphics[width=70mm]{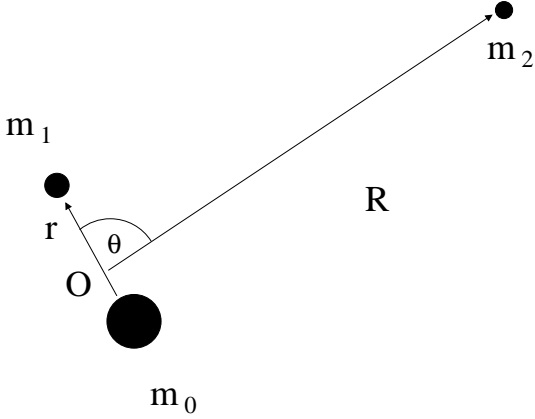}
\caption{The Jacobi decomposition of the three body problem. Note that the relative size of the planetary masses is indicative of the outer case, but the relative configuration is valid for both cases.\label{fig1}}
\end{figure}
\end{center}

Following \cite{Lee-et-al-2003}, we make use of the following variables, which is a set of canonical ones \citep{1961mcm..book.....B,1999ssd..book.....M} and for a coplanar three body system they are defined as:
\begin{eqnarray}
\begin{split}
L_1=&mn_1a^{2}_{1}, \hspace{1cm} &l_1,\\
L_2=&\mathcal Mn_2a^{2}_{2}, \hspace{1cm} &l_2,\\
G_1=&L_{1}\sqrt{1-e^2_1}, \hspace{1cm} &\varpi_1,\\
G_2=&L_{2}\sqrt{1-e^2_2}, \hspace{1cm} &\varpi_2,\label{eq:delaunay}
\end{split}
\end{eqnarray}
where $a_i$, $e_i$, $\varpi_i$ and ${l_i}$ are the semi major axes, the eccentricities, the longitudes of pericentre and mean anomalies of the inner ($i=1$) and the outer ($i=2$) orbit respectively.
Furthermore,  
\begin{displaymath}
m=\frac{m_0m_1}{m_0+m_1}\hspace{0.3cm} \mbox{and} \hspace{0.3cm}\mathcal M =\frac{m_2(m_0+m_1)}{M}, 
\end{displaymath}
are the so-called reduced masses, with $m_0, m_1$ and $m_2$, being the masses of the star and two planets respectively. The total mass of the system is $M=\sum_{i=0}^2m_i$.
Using these variables, the Hamiltonian for a hierarchical triple system reads 
\begin{equation}
\label{ham}
H=H_0+H_1+H_{p},
\end{equation}
where
\begin{equation}
H_0=-\frac{\mathcal{G}^2m^3_0m^3_1}{2(m_0+m_1)L^{2}_{1}}
\end{equation}
is the Keplerian energy of the inner orbit,
\begin{equation}
H_1=-\frac{\mathcal{G}^2(m_0+m_1)^3m^3_2}{2ML^{2}_{2}}
\end{equation}
is the Keplerian energy of the outer orbit, and
\begin{equation}
\label{hp}
H_{p}=\mathcal{G}m_2(\frac{m_0+m_1}{R}-\frac{m_0}{r_{02}}-\frac{m_1}{r_{12}}), 
\end{equation}
is the perturbing Hamiltonian, with ${r_{02}}$ and ${r_{12}}$ being the distances between ${m_{0}}$ and ${m_{2}}$
and  ${m_{1}}$ and ${m_{2}}$ respectively.  $\mathcal{G}$ is the gravitational constant, $\textit{\textbf{R}}$ is the vector which connects the centre of mass of the inner binary with the third body, while $\textit{\textbf{r}}$ will be denoting the relative position vector of the inner orbit (see Figure \ref{fig1}).
Equation (\ref{hp}) can be written as
\begin{equation}
\label{pham}
H_{p}=\mathcal{G}m_{2}(\frac{m_{0}+m_{1}}{R}-\frac{m_{0}}{|\textit{\textbf{R}}+\mu_{1}\textit{\textbf{r}}|}-\frac{m_{1}}{|\textit{\textbf{R}}-\mu_{0}\textit{\textbf{r}}|}),
\end{equation}
with 
\begin{displaymath}
\mu_i=\frac{m_i}{m_0+m_1},\qquad i=0,1. 
\end{displaymath}
Since we deal with a hierarchical system and ${r/R}<<1$, the inverse distances can be presented in terms of Legendre polynomials $P_n$ as follows \citep[e.g. see][]{1999ssd..book.....M}:
\begin{equation}
\frac{m_{0}}{|\textit{\textbf{R}}+\mu_{1}\textit{\textbf{r}}|}=\frac{m_{0}}{R}\sum^{\infty}_{n=0}\left(-\frac{\mu_{1}r}{R}\right)^{n}P_{n}(\cos{\theta})
\end{equation}
and
\begin{equation}
\frac{m_{1}}{|\textit{\textbf{R}}-\mu_{0}\textit{\textbf{r}}|}=\frac{m_{1}}{R}\sum^{\infty}_{n=0}\left(\frac{\mu_{0}r}{R}\right)^{n}P_{n}(\cos{\theta}),
\end{equation}
with $\theta$ being the angle between the $\textit{\textbf{r}}$ and $\textit{\textbf{R}}$ vectors. Then, after some basic algebraic manipulations, equation (\ref{pham}) becomes:
\begin{equation}
H_p=-\frac{\mathcal{G}m_0m_1m_2}{R}\sum^{\infty}_{j=2}M_j(\frac{r}{R})^j\mathcal{P}_j(\cos\theta),
\end{equation}
where 
\begin{displaymath}
M_j=\frac{m^{j-1}_{0}-(-m_1)^{j-1}}{(m_0+m_1)^j}.
\end{displaymath}

In order to describe the long term motion of the system, the above Hamiltonian must be averaged over the fast angles, after suitable expressions for the two relative position vectors and the angle $\theta$ are inserted.  The Legendre polynomials are given by
\begin{equation}  
P_n(\cos{\theta})=2^n\sum_{k=0}^{n} \cos^k{\theta}\binom{n}{k}\binom{\frac{n+k-1}{2}}{n},
\end{equation}
while for the rest of the quantities we use the following expressions:
\begin{eqnarray}
\frac{r}{R} &=& a_1(1-e_1\cos{E_1})\frac{1+e_2\cos{f_2}}{a_2(1-e_2^2)}\\
\cos{\theta} &=& \cos{(f_1+\varpi_1-f_2-\varpi_2)}\\
\cos{f_1} &=& \frac{\cos{E_1}-e_1}{1-e_1\cos{E_1}}\\
\sin{f_1} &=& \sqrt{1-e_1^2}\frac{\sin{E_1}}{1-e_1\cos{E_1}},
\end{eqnarray}
where $E_1$ is the eccentric anomaly of the inner orbit and $f_i$ denotes the true anomaly ($i=1,2$).
In order to find the secular Hamiltonian, we need to filter out the effects with timescales 
of the order of the two planetary periods. Those effects are represented in our Hamiltonian by the inner eccentric anomaly $E_1$ and the outer true anomaly $f_2$ and hence, the secular Hamiltonian can be obtained by calculating the following double integral:
\begin{equation}
\label{aver}
H_l=H_{0l}+H_{1l}+\frac{1}{4\pi^2}\int_0^{2\pi}\int_0^{2\pi} \! (1-e_1\cos{E_1})\frac{(1-e_2^2)^{3/2}}{(1+e_2\cos{f_2})^2}H_p \, \mathrm{d}E_1 \mathrm{d}f_2,
\end{equation}
where the index l denotes long term evolution.  Although for simplicity reasons we will not be using the index l hereinafter,  the Hamiltonian and orbital elements will refer to the long term ones.  

Within this framework, the averaged Hamiltonian can be derived up to any order we may choose. For this work, we have chosen to calculate the Hamiltonian up to the eleventh order. Our main aim is to describe the motion of a terrestrial planet under the gravitational influence of a giant planet when the semi-major axis ratio is rather small ($a_2/a_1$ as low as ~1.6, 1.7). Numerical tests, up to the fifteenth order, have shown negligible differences with the eleventh order expansion for that kind of semi-major axis ratio, suggesting that the latter provides a sufficiently good analytical estimate for the semi-major axis and mass ratios investigated in this work.  However, the analytical estimates are presented in such a way that they can be truncated to any order below the eleventh.  Further justification for this choice is given in section 3.

Since the smaller planet's orbit is initially circular and it is assumed that it is not going to become very eccentric, instead of deriving the complete 
Hamiltonian and then neglecting terms of $O(e^2_i)$ in the secular equations of motion as was done in \cite{Georgakarakos-2003}, we 
choose to make an equivalent approximation by deriving an approximate secular Hamiltonian where terms of order $O(e^3_i)$ and higher are neglected.  The  equations of motion 
based on the approximate Hamiltonian will eventually lead to the same solution we would get if we removed the second order terms in the equations of motion.  However, this way we avoid lengthy expressions for the Hamiltonian and the subsequent equations of motion while maintaining the accuracy of our analytical estimates. The complete (in eccentricities) eleventh order perturbing secular Hamiltonian, which will be used for testing purposes in section 3, can be found in Appendix \ref{ham11} and it is in good agreement with the results obtained by other authors \citep[e.g.][]{2008MNRAS.388..789M,2010A&A...522A..60L}.

\subsection{Inner case}
We begin by deriving the analytical solution for the inner case where the smaller planet's orbit is inside the orbit of the larger planetary body.  From equation (\ref{aver}) and expressed in orbital elements up to the eleventh order, the approximate long term perturbing Hamiltonian becomes :
\begin{equation}
H_{p}=H_{p_2}+H_{p_3}+H_{p_4}+H_{p_5}+H_{p_6}+H_{p_7}+H_{p_8}+H_{p_9}+H_{p_{10}}+H_{p_{11}},
\end{equation}
with 
\begin{eqnarray}
H_{p_2} &=& -\frac{1}{4}\frac{\mathcal{G}m_0m_1m_2}{(m_0+m_1)(1-e_{2}^2)^{3/2}}\frac{a_{1}^2}{a_{2}^3}(1+\frac{3}{2}e_{1}^2)\\
H_{p_3} &=& \frac{15}{16}\frac{\mathcal{G}m_0m_1m_2(m_0-m_1)}{(m_0+m_1)^2(1-e_{2}^2)^{5/2}}\frac{a_{1}^3}{a_{2}^4}e_{1}e_{2}\cos{(\varpi_{1}-\varpi_{2})}\\
H_{p_4} &=& -\frac{9}{64}\frac{\mathcal{G}m_0m_1m_2(m_0^3+m_1^3)}{(m_0+m_1)^4(1-e_{2}^2)^{7/2}}\frac{a_{1}^4}{a_{2}^5}[1+\frac{3}{2}e_{2}^2+(5+\frac{15}{2}e_{2}^2)e_{1}^2+\frac{35}{4}e_{1}^2e_{2}^2\cos{(2\varpi_{1}-2\varpi_{2})}]\\
H_{p_5} &=& \frac{105}{64}\frac{\mathcal{G}m_0m_1m_2(m_0^4-m_1^4)}{(m_0+m_1)^5(1-e_{2}^2)^{9/2}}\frac{a_{1}^5}{a_{2}^6}(1+\frac{3}{4}e_{2}^2)e_{1}e_{2}\cos{(\varpi_{1}-\varpi_{2})}\\
H_{p_6} &=& -\frac{25}{256}\frac{\mathcal{G}m_0m_1m_2(m_0^5+m_1^5)}{(m_0+m_1)^6(1-e_{2}^2)^{11/2}}\frac{a_{1}^6}{a_{2}^7}[1+5e_{2}^2+\frac{15}{8}e_{2}^4+(\frac{21}{2}+\frac{105}{2}e_{2}^2+\frac{315}{16}e_{2}^4)e_{1}^2+(\frac{189}{4}+\frac{189}{8}e_{2}^2)e_{1}^2e_{2}^2\times\nonumber\\
& & \cos{(2\varpi_{1}-2\varpi_{2})}]\\
H_{p_7} &=& \frac{4725}{2048}\frac{\mathcal{G}m_0m_1m_2(m_0^6-m_1^6)}{(m_0+m_1)^7(1-e_{2}^2)^{13/2}}\frac{a_{1}^7}{a_{2}^8}(1+\frac{5}{2}e_{2}^2+\frac{5}{8}e_{2}^4)e_{1}e_{2}\cos{(\varpi_{1}-\varpi_{2})}\\
H_{p_8} &=& -\frac{1225}{8192}\frac{\mathcal{G}m_0m_1m_2(m_0^7+m_1^7)}{(m_0+m_1)^8(1-e_{2}^2)^{15/2}}\frac{a_{1}^8}{a_{2}^9}[\frac{1}{2}+\frac{21}{4}e_{2}^2+\frac{105}{16}e_{2}^4+\frac{35}{32}e_{2}^6+(9+\frac{189}{2}e_{2}^2+\frac{945}{8}e_{2}^4+\frac{315}{16}e_{2}^6)e_{1}^2+\nonumber\\
& & (\frac{297}{4}+\frac{495}{4}e_{2}^2+\frac{1485}{64}e_{2}^4)e_{1}^2e_{2}^2\cos{(2\varpi_{1}-2\varpi_{2})}]\\
H_{p_9} &=& \frac{24255}{8192}\frac{\mathcal{G}m_0m_1m_2(m_0^8-m_1^8)}{(m_0+m_1)^9(1-e_{2}^2)^{17/2}}\frac{a_{1}^9}{a_{2}^{10}}(1+\frac{21}{4}e_{2}^2+\frac{35}{8}e_{2}^4+\frac{35}{64}e_{2}^6)e_{1}e_{2}\cos{(\varpi_{1}-\varpi_{2})}\\
H_{p_{10}} &=& -\frac{3969}{32768}\frac{\mathcal{G}m_0m_1m_2(m_0^9+m_1^9)}{(m_0+m_1)^{10}(1-e_{2}^2)^{19/2}}\frac{a_{1}^{10}}{a_{2}^{11}} 
[\frac{1}{2}+9e_{2}^2+\frac{189}{8}e_{2}^4+\frac{105}{8}e_{2}^6+\frac{315}{256}e_{2}^8+
(\frac{55}{4}+\frac{495}{2}e_{2}^2+\frac{10395}{16}e_{2}^4+\nonumber\\
& & +\frac{5775}{16}e_{2}^6+\frac{17325}{512}e_{2}^8)e_{1}^2+(\frac{715}{4}+\frac{5005}{8}e_{2}^2+\frac{25025}{64}e_{2}^4+\frac{5005}{128}e_{2}^6)
e_{1}^2e_{2}^2\cos{(2\varpi_{1}-2\varpi_{2})}]\\
H_{p_{11}} &=& \frac{455}{262144}\frac{\mathcal{G}m_0m_1m_2(m_0^{10}-m_1^{10})}{(m_0+m_1)^{11}(1-e_{2}^2)^{21/2}}\frac{a_{1}^{11}}{a_{2}^{12}}(2079+18711e_{2}^2+\frac{130977}{4}e_{2}^4+\frac{218295}{16}e_{2}^6
+\frac{130977}{128}e_{2}^8)e_{1}e_{2}\times\nonumber\\
& &\times\cos{(\varpi_{1}-\varpi_{2})}
\end{eqnarray}

Having found our Hamiltonian, we are now able to derive the equations of motion of our three body system.  
Using our set of canonical variables, Hamilton's equations assume the following form for a coplanar system:
\begin{eqnarray}
\begin{split}
\frac{dL_{i}}{dt}=-\frac{\partial H}{\partial l_{i}},& \hspace{0.5cm} \frac{dl_{i}}{dt}=\frac{\partial H}{\partial L_{i}}\\
\frac{dG_{i}}{dt}=-\frac{\partial H}{\partial \varpi_{i}},& \hspace{0.5cm} \frac{d\varpi_{i}}{dt}=\frac{\partial H}{\partial G_{i}}, \hspace{0.5cm}i=1,2.
\end{split}
\end{eqnarray}

Since the usual orbital elements are not a set of canonical variables, we can use the above relations to obtain our equations of motion in terms of orbital elements (more details in appendix \ref{A1}). Keeping in mind that we have assumed that the orbit of the larger planet is unaffected by the presence of the smaller planet and that there is no secular change in the semi-major axes (since the Hamiltonian is independent of $l_i$), we shall only need expressions for the eccentricity and longitude of pericentre of the orbit of the smaller planet.  Those are:

\begin{eqnarray}
\frac{de_{1}}{dt} &=&\frac{(m_0+m_1)^{1/2}(1-e_{1}^2)^{1/2}}{\mathcal{G}^{1/2}m_0m_1a_{1}^{1/2}e_{1}}\frac{\partial H}{\partial \varpi_{1}}\\
\frac{d\varpi_{1}}{dt} &=&-\frac{(m_0+m_1)^{1/2}(1-e_{1}^2)^{1/2}}{\mathcal{G}^{1/2}m_0m_1a_{1}^{1/2}e_{1}}\frac{\partial H}{\partial e_{1}}.
\end{eqnarray}

However, instead of using $e_{1}$ and $\varpi_{1}$, it is preferable to use the variables $x_{1}=e_{1}\cos{\varpi_{1}}$ and $y_{1}=e_{1}\sin{\varpi_{1}}$, i.e. the components of the long term inner eccentric vector. In this way we can avoid singularities in the equations of motion when the eccentricity of the smaller body is initially zero.  Using the expressions for the long term Hamiltonian we derived earlier, we get the following system of differential equations:

\begin{eqnarray}
\label{de1}
\dot{x}_{1}&=& A\sin{2\varpi_{2}}x_{1}+(B-2A\cos^2{\varpi_{2}})y_{1}+C\sin{\varpi_{2}}\\
\label{de2}
\dot{y}_{1}&=& -(B-2A\sin^2{\varpi_{2}})x_{1}-A\sin{2\varpi_{2}}y_{1}-C\cos{\varpi_{2}},
\end{eqnarray}
where $A, B$ and $C$ are constants and are given by:

\begin{equation}
A=A_{p4}+A_{p6}+A_{p8}+A_{p10}
\end{equation}
\begin{eqnarray}
A_{p4} &=&-\frac{315}{128}\frac{\sqrt{G}m_3(m^3_1+m^3_2)}{(m_1+m_2)^{7/2}(1-e^2_{2})^{7/2}}\frac{a^{7/2}_{1}}{a^5_{2}}e^2_{2}\\
A_{p6} &=& -\frac{4725}{1024}\frac{\sqrt{G}m_3(m^5_1+m^5_2)}{(m_1+m_2)^{11/2}(1-e^2_{2})^{11/2}}\frac{a^{11/2}_{1}}{a^7_{2}}e_{2}^2(2+e^2_{2})\\
A_{p8} &=& -\frac{11025}{262144}\frac{\sqrt{G}m_3(m^7_1+m^7_2)}{(m_1+m_2)^{15/2}(1-e^2_{2})^{15/2}}\frac{a^{15/2}_{1}}{a^9_{2}}e^2_{2}(528+880e^2_{2}+165e^4_{2})\\
A_{p10} &=& -\frac{218295}{8388608}\frac{\sqrt{G}m_3(m^9_1+m^9_2)}{(m_1+m_2)^{19/2}(1-e^2_{2})^{19/2}}\frac{a^{19/2}_{1}}{a^{11}_{2}}e^2_{2}(1664+5824e^2_{2}+3640e^4_{2}+364e^6_{2})
\end{eqnarray}
\begin{equation}
B=B_{p2}+B_{p4}+B_{p6}+B_{p8}+B_{p10}
\end{equation}
\begin{eqnarray}
B_{p2} &=& -\frac{3}{4}\frac{\sqrt{G}m_3}{(m_1+m_2)^{1/2}(1-e^2_2)^{3/2}}\frac{a^{3/2}_{1}}{a^3_{2}}\\
B_{p4} &=& -\frac{45}{128}\frac{\sqrt{G}m_3(m^3_1+m^3_2)}{(m_1+m_2)^{7/2}(1-e^2_{2})^{7/2}}\frac{a^{7/2}_{1}}{a^{5}_{2}}(4+13e^2_{2})\\
B_{p6} &=& -\frac{525}{2048}\frac{\sqrt{G}m_3(m^5_1+m^5_2)}{(m_1+m_2)^{11/2}(1-e^2_{2})^{11/2}}\frac{a^{11/2}_{1}}{a^{7}_{2}}(8+76e^2_{2}+33e^4_{2})\\
B_{p8} &=& -\frac{11025}{262144}\frac{\sqrt{G}m_3(m^7_1+m^7_2)}{(m_1+m_2)^{15/2}(1-e^2_2)^{15/2}}\frac{a^{15/2}_1}{a^{9}_2}(64+1200e^2_2+1720e^4_2+305e^6_2)\\
B_{p10} &=& -\frac{218295}{8388608}\frac{\sqrt{G}m_3(m^9_1+m^9_2)}{(m_1+m_2)^{19/2}(1-e^2_{2})^{19/2}}\frac{a^{19/2}_{1}}{a^{11}_{2}}(128+3968e^2_{2}+11872e^4_{2}+7000e^6_{2}+679e^8_{2})\\
\end{eqnarray}
\begin{equation}
C=C_{p3}+C_{p5}+C_{p7}+C_{p9}+C_{p11}
\end{equation}
\begin{eqnarray}
C_{p3}&=&\frac{15}{16}\frac{\sqrt{G}m_3(m_1-m_2)}{(m_1+m_2)^{3/2}(1-e^2_{2})^{5/2}}\frac{a^{5/2}_{1}}{a^{4}_{2}}e_{2}\\
C_{p5}&=&\frac{105}{256}\frac{\sqrt{G}m_3(m^4_1-m^4_2)}{(m_1+m_2)^{9/2}(1-e^2_{2})^{9/2}}\frac{a^{9/2}_{1}}{a^{6}_{2}}(4+3e^2_{2})e_{2}\\
C_{p7}&=&\frac{4725}{16384}\frac{\sqrt{G}m_3(m^6_1-m^6_2)}{(m_1+m_2)^{13/2}(1-e^2_{2})^{13/2}}\frac{a^{13/2}_{1}}{a^{8}_2}(8+20e^2_{2}+5e^4_{2})e_{2}\\
C_{p9}&=&\frac{24255}{524288}\frac{\sqrt{G}m_3(m^8_1-m^8_2)}{(m_1+m_2)^{17/2}(1-e^2_{2})^{17/2}}\frac{a^{17/2}_{1}}{a^{10}_{2}}(64+336e^2_{2}+280e^4_{2}+35e^6_{2})e_{2}\\
C_{p11}&=&\frac{945945}{33554432}\frac{\sqrt{G}m_3(m^{10}_1-m^{10}_2)}{(m_1+m_2)^{21/2}(1-e^2_{2})^{21/2}}\frac{a^{21/2}_{1}}{a^{12}_{2}}(128+1152e^2_{2}+2016e^4_{2}+840e^6_{2}+63e^8_{2})e_{2}.
\end{eqnarray}

The solution to equations (\ref{de1}) and (\ref{de2}) is
\begin{eqnarray}
\label{inner1}
x_{1} &=& C_1\cos{\sqrt{B^2-2AB}t}+C_2\sin{\sqrt{B^2-2AB}t}-\frac{C}{B}\cos{\varpi_{2}}\\
\label{inner2}
y_{1}&=&\frac{1}{B-2A\cos^2{\varpi_{2}}}[(C_2\sqrt{B^2-2AB}-AC_1\sin{2\varpi_{2}})\cos{\sqrt{B^2-2AB}t}-(C_1\sqrt{B^2-2AB}+\nonumber\\& & +AC_2\sin{2\varpi_{2}})\sin{\sqrt{B^2-2AB}t}]-\frac{C}{B}\sin{\varpi_{2}},
\end{eqnarray}
where $C_1$ and $C_2$ are constants of integrations. Since the components of the eccentric vector are initially zero, i.e. $x_{10}=0$ and  $y_{10}=0$, the constants $C_1$ and $C_2$ are given by
\begin{eqnarray}
\label{const1}
C_1 &=& \frac{C}{B}\cos{\varpi_{2}}\\
\label{const2}
C_2 &=& \frac{C}{\sqrt{B(B-2A)}}\sin{\varpi_{2}}.
\end{eqnarray}
If we substitute the above values for our constants into equations (\ref{inner1}) and (\ref{inner2}), the latter become:
\begin{eqnarray}
\label{finner1}
x_{1} &=& \frac{C}{B}\cos{\varpi_{2}}\cos{\sqrt{B^2-2AB}t}+\frac{C}{\sqrt{B^2-2AB}}\sin{\varpi_{2}}\sin{\sqrt{B^2-2AB}t}-\frac{C}{B}\cos{\varpi_{2}}\\
\label{finner2}
y_{1} &=& \frac{C}{B}\sin{\varpi_{2}}\cos{\sqrt{B^2-2AB}t}-\frac{C}{\sqrt{B^2-2AB}}\cos{\varpi_{2}}\sin{\sqrt{B^2-2AB}t}-\frac{C}{B}\sin{\varpi_{2}}.
\end{eqnarray}
Hence,
\begin{eqnarray}
e_{1}&=&\sqrt{x^2_{1}+y^2_{1}}\\
\varpi_{1}&=&\arctan{(\frac{y_{1}}{x_{1}})}.
\end{eqnarray}

\subsection{Outer case}
Similarly to the inner case, we can derive our long term Hamiltonian for the case where the giant planet is inside the orbit of the smaller planet.  The Hamiltonian in this case is obtained in a similar fashion to the one for the inner case; the only difference is that we now neglect terms of $O(e^3_{2})$ and higher instead of terms of at least of order $O(e^3_{1})$.  As a result, the long term perturbing Hamiltonian is:

\begin{eqnarray}
H_{p_2} &=& -\frac{1}{4}\frac{\mathcal{G}m_0m_1m_2}{(m_0+m_1)(1-e_{2}^2)^{3/2}}\frac{a_{1}^2}{a_{2}^3}(1+\frac{3}{2}e_{1}^2)\\
H_{p_3} &=& \frac{15}{16}\frac{\mathcal{G}m_0m_1m_2(m_0-m_1)}{(m_0+m_1)^2(1-e_{2}^2)^{5/2}}\frac{a_{1}^3}{a_{2}^4}e_{1}e_{2}(1+\frac{3}{4}e_{1}^2)\cos{(\varpi_{1}-\varpi_{2})}\\
H_{p_{4}} &=& -\frac{9}{64}\frac{\mathcal{G}m_0m_1m_2(m_0^3+m_1^3)}{(m_0+m_1)^4(1-e_{2}^2)^{7/2}}\frac{a_{1}^4}{a_{2}^5}[1+5e_{1}^2+\frac{15}{8}e_{1}^4+(\frac{3}{2}+\frac{15}{2}e_{1}^2+\frac{45}{16}e_{1}^4)e_{2}^2+(\frac{35}{4}+\frac{35}{8}e_{1}^2)e_{1}^2e_{2}^2\times\nonumber\\
& & \times\cos{(2\varpi_{1}-2\varpi_{2})}]\\
H_{p_{5}} &=& \frac{105}{64}\frac{\mathcal{G}m_0m_1m_2(m_0^4-m_1^4)}{(m_0+m_1)^5(1-e_{2}^2)^{9/2}}\frac{a_{1}^5}{a_{2}^6}(1+\frac{5}{2}e_{1}^2+\frac{5}{8}e_{1}^4)e_{1}e_{2}\cos{(\varpi_{1}-\varpi_{2})}\\
H_{p_{6}} &=& -\frac{25}{256}\frac{\mathcal{G}m_0m_1m_2(m_0^5+m_1^5)}{(m_0+m_1)^6(1-e_{2}^2)^{11/2}}\frac{a_{1}^6}{a_{2}^7}[1+\frac{21}{2}e_{1}^2+\frac{105}{8}e_{1}^4+\frac{35}{16}e_{1}^6+(5+\frac{105}{2}e_{1}^2+\frac{525}{8}e_{1}^4+\nonumber\\
& & +\frac{175}{16}e_{1}^6)e_{2}^2+(\frac{189}{4}+\frac{315}{4}e_{1}^2+\frac{945}{64}e_{1}^4)e_{1}^2e_{2}^2\cos{(2\varpi_{1}-2\varpi_{2})}]\\
H_{p_{7}} &=& \frac{4725}{2048}\frac{\mathcal{G}m_0m_1m_2(m_0^6-m_1^6)}{(m_0+m_1)^7(1-e_{2}^2)^{13/2}}\frac{a_{1}^7}{a_{2}^8}
(1+\frac{21}{4}e_{1}^2+\frac{35}{8}e_{1}^4+\frac{35}{64}e_{1}^6)e_{1}e_{2}\cos{(\varpi_{1}-\varpi_{2})}\\
H_{p_{8}} &=& -\frac{1225}{8192}\frac{\mathcal{G}m_0m_1m_2(m_0^7+m_1^7)}{(m_0+m_1)^8(1-e_{2}^2)^{15/2}}\frac{a_{1}^8}{a_{2}^9}[\frac{1}{2}+9e_{1}^2+\frac{189}{8}e_{1}^4+\frac{105}{8}e_{1}^6+\frac{315}{256}e_{1}^8+
(\frac{21}{4}+\nonumber\\
& & +\frac{189}{2}e_{1}^2+\frac{3969}{16}e_{1}^4+\frac{2205}{16}e_{1}^6+\frac{6615}{512}e_{1}^8)e_{2}^2+(\frac{297}{4}+\frac{2079}{8}e_{1}^2+\frac{10395}{64}e_{1}^4+\nonumber\\
& & +\frac{2079}{128}e_{1}^6)
e_{1}^2e_{2}^2\cos{(2\varpi_{1}-2\varpi_{2})}]\\
H_{p_{9}} &=& \frac{24255}{8192}\frac{\mathcal{G}m_0m_1m_2(m_0^8-m_1^8)}{(m_0+m_1)^9(1-e_{2}^2)^{17/2}}\frac{a_{1}^9}{a_{2}^{10}} 
(1+9e_{1}^2+\frac{63}{4}e_{1}^4+\frac{105}{16}e_{1}^6)e_{1}e_{2}\cos{(\varpi_{1}-\varpi_{2})}\\
H_{p_{10}} &=& -\frac{3969}{32768}\frac{\mathcal{G}m_0m_1m_2(m_0^9+m_1^9)}{(m_0+m_1)^{10}(1-e_{2}^2)^{19/2}}\frac{a_{1}^{10}}{a_{2}^{11}} [\frac{1}{2}+\frac{55}{4}e_{1}^2+\frac{495}{8}e_{1}^4+\frac{1155}{16}e_{1}^6+\frac{5775}{256}e_{1}^8+(9+\nonumber\\
& & +\frac{495}{2}e_{1}^2+\frac{4455}{4}e_{1}^4+\frac{10395}{8}e_{1}^6+\frac{51975}{128}e_{1}^8)e_{2}^2+(\frac{715}{4}+\frac{2145}{2}e_{1}^2+\frac{45045}{32}e_{1}^4+\nonumber\\
& & +\frac{15015}{32}e_{1}^6)
e_{1}^2e_{2}^2\cos{(2\varpi_{1}-2\varpi_{2})}]\\
H_{p_{11}}&=& \frac{945945}{262144}\frac{\mathcal{G}m_0m_1m_2(m_0^{10}-m_1^{10})}{(m_0+m_1)^{11}(1-e_{2}^2)^{21/2}}\frac{a_{1}^{11}}{a_{2}^{12}}(1+\frac{55}{4}e_{1}^2+\frac{165}{4}e_{1}^4+\frac{1155}{32}e_{1}^6)e_{1}e_{2}\cos{(\varpi_{1}-\varpi_{2})}.
\end{eqnarray}
Note that $H_{p_2}$ is the same as in the inner case.

Following the same procedure as before, we end up again with a system of two linear non-homogeneous differential equations,
which has the same form as the inner case:

\begin{eqnarray}
\label{de3}
\dot{x}_{2} &=& A\sin{2\varpi_{1}}x_{2}+(B-2A\cos^2{\varpi_{1}})y_{2}+C\sin{\varpi_{1}}\\
\label{de4}
\dot{y}_{2} &=& -(B-2A\sin^2{\varpi_{1}})x_{2}-A\sin{2\varpi_1}y_{2}-C\cos{\varpi_{1}},
\end{eqnarray}

with
\begin{equation}
A=A_{p4}+A_{p6}+A_{p8}+A_{p10}
\end{equation}
\begin{eqnarray}
A_{p4} &=&-\frac{315}{256}\frac{\sqrt{GM}m_1m_2(m^3_1+m^3_2)}{(m_1+m_2)^5}\frac{a^4_{1}}{a^{11/2}_{2}}(2+e^2_{1})e^2_{1}\\
A_{p6} &=&-\frac{75}{8192}\frac{\sqrt{GM}m_1m_2(m^5_1+m^5_2)}{(m_1+m_2)^7}\frac{a^6_{1}}{a^{15/2}_{2}}(1008+1680e^2_{1}+315e^4_{1})e^2_{1}\\
A_{p8} &=&-\frac{3675}{524288}\frac{\sqrt{GM}m_1m_2(m^7_1+m^7_2)}{(m_1+m_2)^9}\frac{a^8_{1}}{a^{19/2}_{2}}(3168+11088e^2_{1}+6930e^4_{1}+693e^6_{1})e^2_{1}\\
A_{p10} &=&-\frac{3969}{8388608}\frac{\sqrt{GM}m_1m_2(m^9_1+m^9_2)}{(m_1+m_2)^{11}}\frac{a^{10}_{1}}{a^{23/2}_{2}}(91520+549120e^2_{1}+720720e^4_{1}+240240e^6_{1})e^2_{1}
\end{eqnarray}
\begin{equation}
B=B_{p2}+B_{p4}+B_{p6}+B_{p8}+B_{p10}
\end{equation}
\begin{eqnarray}
B_{p2} &=&-\frac{3}{8}\frac{\sqrt{GM}m_1m_2}{(m_1+m_2)^2}\frac{a^2_{1}}{a^{7/2}_{2}}(2+3e^2_{1})\\
B_{p4} &=&-\frac{45}{256}\frac{\sqrt{GM}m_1m_2(m^3_1+m^3_2)}{(m_1+m_2)^5}\frac{a^4_{1}}{a^{11/2}_{2}}
(8+54e^2_{1}+22e^4_{1})\\
B_{p6} &=&-\frac{75}{8192}\frac{\sqrt{GM}m_1m_2(m^5_1+m^5_2)}{(m_1+m_2)^7}\frac{a^6_{1}}{a^{15/2}_{2}}
(224+3360e^2_{1}+4620e^4_{1}+805e^6_{1})\\
B_{p8} &=& -\frac{3675}{524288}\frac{\sqrt{GM}m_1m_2(m^7_1+m^7_2)}{(m_1+m_2)^9}\frac{a^8_{1}}{a^{19/2}_{2}}            
(384+10080e^2_{1}+292332e^4_{1}+17010e^6_{1}+1638e^8_{1})\\
B_{p10} &=& -\frac{3969}{8388608}\frac{\sqrt{GM}m_1m_2(m^9_1+m^9_2)}{(m_1+m_2)^{11}}\frac{a^{10}_{1}}{a^{23/2}_{2}}(7040+285120e^2_{1}+1420320e^4_{1}+1737120e^6_{1}+557865e^8_{1})
\end{eqnarray}

\begin{equation}
C=C_{p3}+C_{p5}+C_{p7}+C_{p9}+C_{p11}
\end{equation}
\begin{eqnarray}
C_{p3}&=&\frac{15}{64}\frac{\sqrt{GM}m_1m_2(m_1-m_2)}{(m_1+m_2)^{3}}\frac{a^{3}_{1}}{a^{9/2}_{2}}(4+3e^2_{1})e_{1}\\
C_{p5}&=&\frac{105}{512}\frac{\sqrt{GM}m_1m_2(m^{4}_1-m^{4}_2)}{(m_1+m_2)^{6}}\frac{a^{5}_{1}}{a^{13/2}_{2}}(8+20e^2_{1}+5e^4_{1})e_{1}\\
C_{p7}&=&\frac{4725}{131072}\frac{\sqrt{GM}m_1m_2(m^{6}_1-m^{6}_2)}{(m_1+m_2)^{8}}\frac{a^{7}_{1}}{a^{17/2}_{2}}(64+336e^2_{1}+280e^4_{1}+35e^6_{1})e_{1}\\
C_{p9}&=&\frac{24255}{131072}\frac{\sqrt{GM}m_1m_2(m^{8}_1-m^{8}_2)}{(m_1+m_2)^{10}}\frac{a^{9}_{1}}{a^{21/2}_{2}}(16+144e^2_{1}+252e^4_{1}+105e^6_{1})e_{1}\\
C_{p11}&=&\frac{945945}{8388608}\frac{\sqrt{GM}m_1m_2(m^{10}_1-m^{10}_2)}{(m_1+m_2)^{12}}\frac{a^{11}_{1}}{a^{25/2}_{2}}(32+440e^2_{1}+1320e^4_{1}+1155e^6_{1})e_{1}.
\end{eqnarray}

The solution to equations (\ref{de3}) and (\ref{de4}) has the same form as the one for the inner case:

\begin{eqnarray}
\label{fouter1}
x_{2} &=& \frac{C}{B}\cos{\varpi_{1}}\cos{\sqrt{B^2-2AB}t}+\frac{C}{\sqrt{B^2-2AB}}\sin{\varpi_{1}}\sin{\sqrt{B^2-2AB}t}-\frac{C}{B}\cos{\varpi_{1}}\\
\label{fouter2}
y_{2} &=& \frac{C}{B}\sin{\varpi_{1}}\cos{\sqrt{B^2-2AB}t}-\frac{C}{\sqrt{B^2-2AB}}\cos{\varpi_{1}}\sin{\sqrt{B^2-2AB}t}-\frac{C}{B}\sin{\varpi_{1}}
\end{eqnarray}
and
\begin{eqnarray}
e_{2}&=&\sqrt{x^2_{2}+y^2_{2}}\\
\varpi_{2}&=&\arctan{(\frac{y_{2}}{x_{2}})}.
\end{eqnarray}

For both the inner and outer cases, the period of the oscillation is 
\begin{eqnarray}
T=\frac{2\pi}{\sqrt{B^2-2AB}}.\label{period}
\end{eqnarray}

\subsection{Maximum and averaged eccentricity}
\label{maxe}
Based on the analytical results of the previous sections, we can find expressions for the maximum
and the averaged square eccentricity.  Using equations (\ref{finner1}), (\ref{finner2}), (\ref{fouter1}) and (\ref{fouter2}) we get: 

\begin{equation}
\label{e2}
e^2_{i}=x^2_{i}+y^2_{i}=\left(\frac{C}{B}\right)^2(1-\cos{\sqrt{B^2-2AB}t})^2+\frac{C^2}{B(B-2A)}\sin^2{\sqrt{B^2-2AB}t}, \hspace{1cm} i=1,2
\end{equation}
As one may see, the above result is independent of the longitude of pericentre of the larger planet.  Thus, averaging the above equation over time, we obtain:

\begin{equation}
\label{ave}
\langle e^2_{i}\rangle =\frac{3}{2}\left(\frac{C}{B}\right)^2+\frac{1}{2}\frac{C^2}{B(B-2A)}=\left(\frac{C}{B}\right)^2\frac{2B-3A}{B-2A} \hspace{1cm} i=1,2.
\end{equation}
In addition, it can easily be shown from equation (\ref{e2}) that the maximum value of the eccentricity is
\begin{equation}
\label{max}
e^{max}_{i}=-2\frac{C}{B} \hspace{1cm} i=1,2.
\end{equation}

\section{Numerical results}

In order to test the validity of our analytical estimates,  we ran a series of numerical experiments which involved integrating both the full equations of motion and the secular equations of motion of the complete Hamiltonian up to the eleventh order.  For the former, we used the symplectic integrator with time transformation developed by Seppo Mikkola \citep{1997CeMDA..67..145M}, which is specially designed to integrate hierarchical triple systems.  The code uses standard Jacobi coordinates, i.e. it calculates the relative position and velocity vectors of the inner and outer orbit at every time step. For the numerical integration of the secular equations, we used the Bulirsch-Stoer integrator given in  \cite{1992nrfa.book.....P}.  

For all simulations the star was one Solar mass, while the mass of the smaller planet was set to one Earth mass. For the larger
planet we used four different masses:\hspace{0.1cm}$m_2=10, 1, 0.1 \hspace{0.2cm}\mbox{and}\hspace{0.2cm} 0.03 M_J$ (which is approximately $10 M_{\oplus}$).  For the inner case the small planet semi-major axis was fixed at 1 au, while the larger planet's semi-major axis ranged from 1.1 au to 7 au and its eccentricity was given values in the interval [0.05,0.4] with a step of 0.05.  Choosing higher values for the
eccentricity of the perturber (>0.4) would be meaningless in our case, since the new analytical estimates are important for smaller
semi-major axis ratios; a system with higher giant planet eccentricity would require greater semi-major axis ratio in 
order to be stable.  In this case, the old solution of \cite{Georgakarakos-2003} with a $P_3$ Legendre polynomial expansion would be sufficient.  In the present study, for any  pair of $(a_2,e_2)$, the initial longitude of pericentre and true anomaly of the larger planet were given the values $0^{\circ}, 90^{\circ},180^{\circ},270^{\circ}$, i.e. we simulated sixteen different orbits. Finally, the smaller planet was always started along the positive direction of the x-axis.
Similarly, for the outer case, we fixed the semi-major axis of the giant planet to 0.5 au, while the smaller planet's semi-major axis was varied from 0.7 au to 3 au. The eccentricity, longitude of pericentre and true anomaly of the giant planet and the initial position of the smaller planet were set in the same way as in the inner case.  For all the simulations the integration time was one analytic secular period as given in equation (\ref{period}).

As we stated in the introduction, this work mainly aims at systems with the two planets relatively close but still on stable orbits \citep[an idea about the stability border for the systems we investigate can be taken from][]{2013NewA...23...41G, 2015ApJ...808..120P}, as our previously developed models are inadequate for such systems.  The quantities we tested were the averaged square eccentricity and the maximum eccentricity.   For both inner and outer case we found the relative error between equation (\ref{ave}) and the averaged square eccentricity from the numerical integration of the full equations of motion.  At the same time,  we measured the same quantity but this time using the 
secular equations of motion derived from the complete (in eccentricities) eleventh order Hamiltonian.  This ensures that a small error between the analytical estimates and the numerical solution of the full problem is not some kind of artifact.  Only a small error in the former accompanied by a small error in
the latter would imply that the analytical estimates were a good approximation to the full problem. We applied the same course of action to the maximum eccentricity obtained for a given
pair of $(a_i,e_i)$, ($i=2,1$ for the inner and outer case respectively).  Finally, for each $(a_i,e_i)$ pair, we recorded the maximum eccentricity the smaller planet achieved during the numerical integration.  

The analytical estimates were generally in good agreement with the results from the numerical simulations.  As expected, the analytical estimates fail when the 
perturbations become strong and the generated eccentricity of the smaller planet reaches moderate to high values or when the system approaches the stability limit.  That normally happens
for higher values of the giant planet eccentricity and smaller semi-major axis ratios.  For the rest of the values in the semi-major axis - eccentricity space, the analytical estimates proved to be a good approximation to the full problem. An exception to that rule were systems that were near mean motion resonances. In such a case, secular
theory fails and the theoretical model developed here does not work well.  The method fails completely near
the exact mean motion resonance location, but gradually improves as we move away from it.  Of course, the extent of the effect depends on the mass of the perturber and its eccentricity, as the width of the resonance depends on those parameters \citep[e.g. see][]{1999ssd..book.....M}. 

\begin{center}
\begin{figure}
\includegraphics[width=92mm,height=120mm]{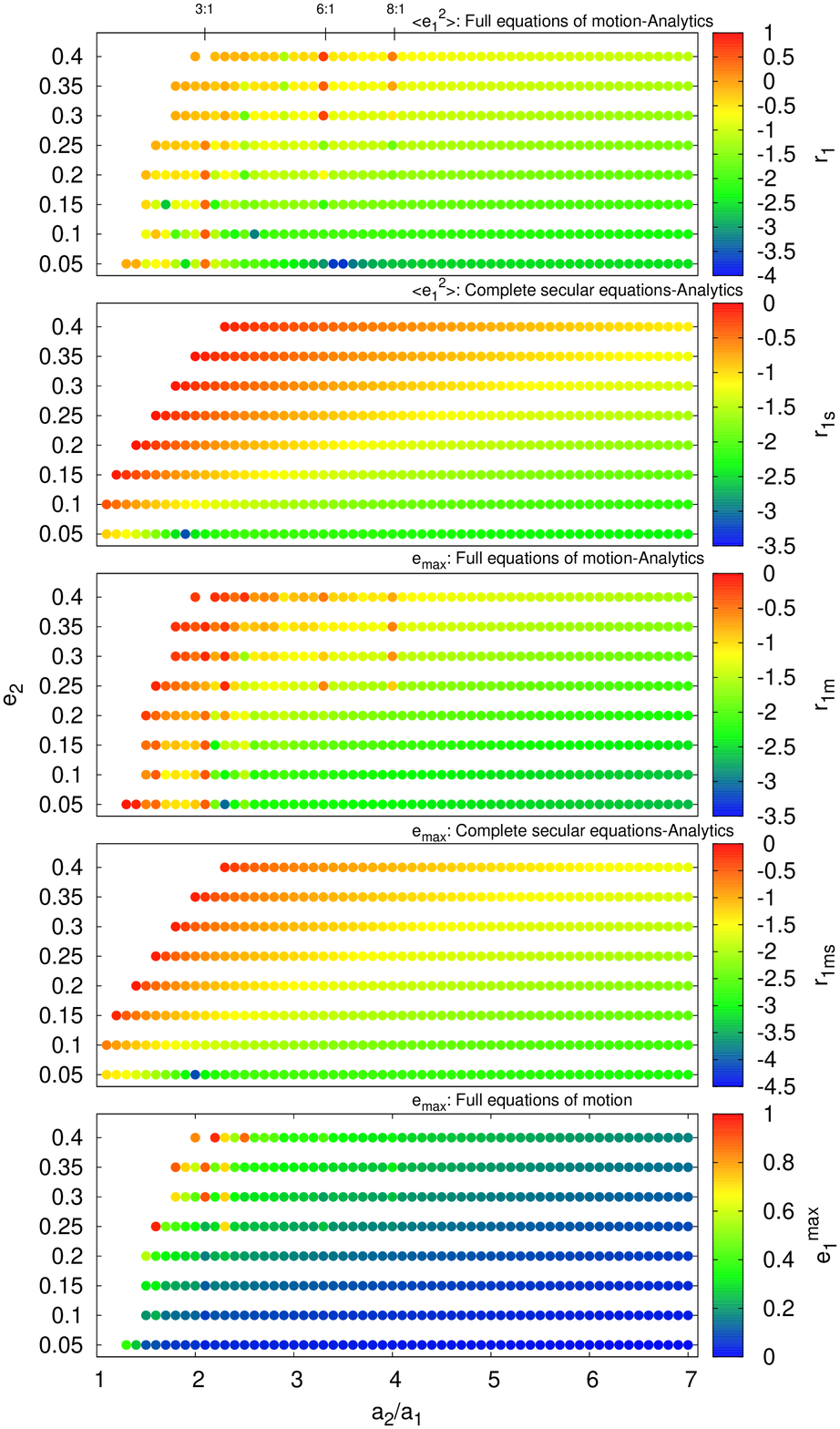}
\includegraphics[width=92mm,height=120mm]{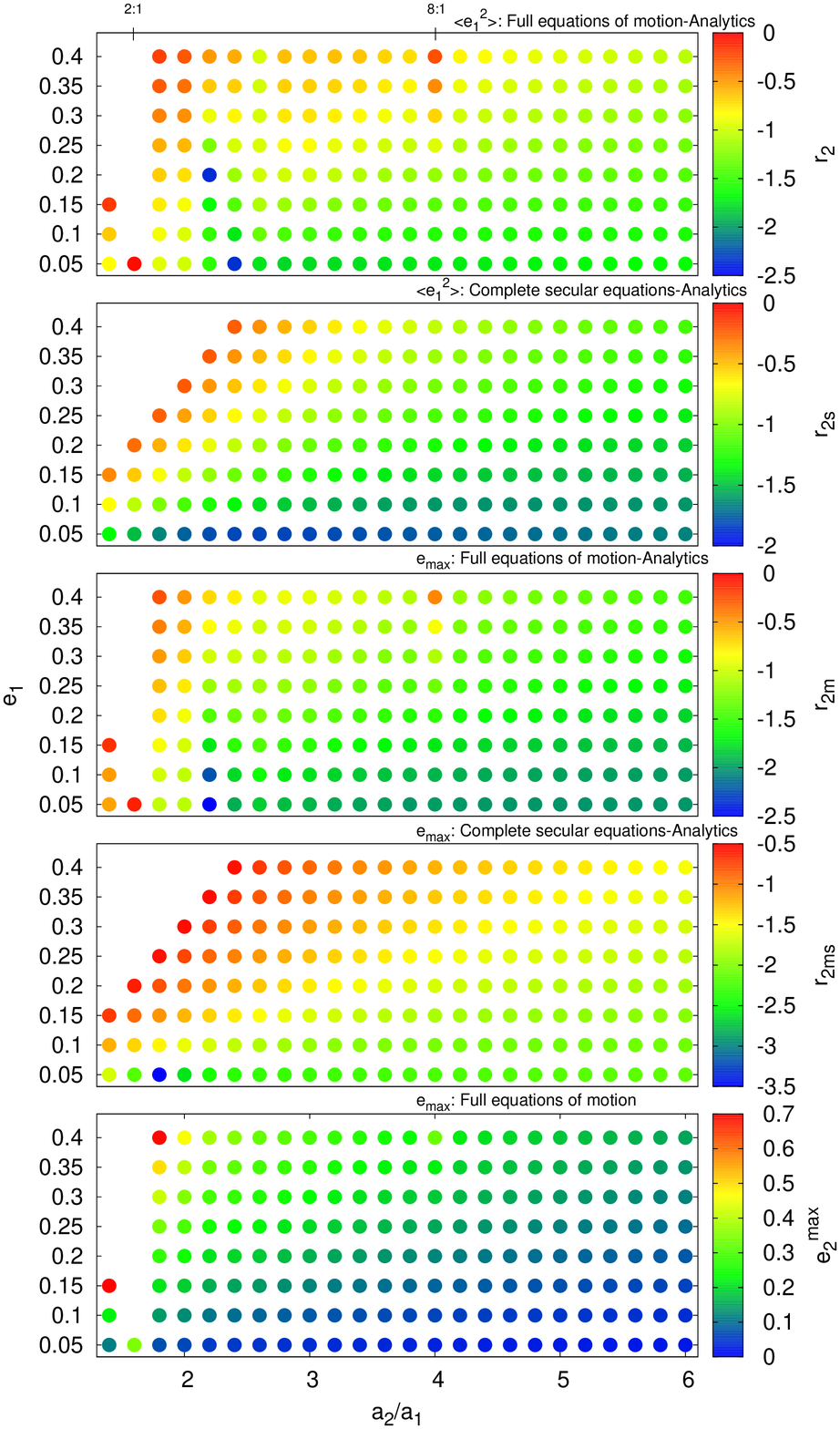}
\caption{Errors for the averaged square and maximum eccentricity for a system with $m_0=1 M_{\odot}, 
m_1=1 M_{\oplus}, m_2=1 M_J$ (left column-inner case) and a system with $m_0=1 M_{\odot}, m_1=1 M_J , m_2=M_{\oplus}$ (right column-outer case). For both columns, from the top: the first row shows the logarithmic relative error in the averaged square eccentricity between the results from the numerical integration of the full equations of motion and our analytical estimates, i.e. $r_i=\log_{10}[(\langle e^2_{in} \rangle-\langle e^2_{i} \rangle)/\langle e^2_{in} \rangle]$.  The second row shows the logarithmic relative error in the averaged square eccentricity between the results from the numerical integration of the full eleventh order secular equations of motion and our analytical estimates, i.e. $r_{is}=\log_{10}[(\langle e^2_{is} \rangle-\langle e^2_{i} \rangle)/\langle e^2_{is} \rangle]$. The third row shows the logarithmic relative error in the maximum eccentricity between the results from the numerical integration of the full equations of motion and our analytical estimates, i.e. $r_{im}=\log_{10}[(\langle e^{max}_{in} \rangle-\langle e^{max}_{i} \rangle)/\langle e^{max}_{in} \rangle]$. The fourth row shows the logarithmic relative error in the maximum eccentricity between the results from the numerical integration of the full eleventh order secular equations of motion and our analytical estimates, i.e. $r_{ims}=\log_{10}[(\langle e^{max}_{is} \rangle-\langle e^{max}_{i} \rangle)/\langle e^{max}_{is} \rangle]$.  Finally, the fifth row shows the maximum eccentricity of the smaller planet $e^{max}_{in}$ achieved during the numerical integration of the full equations of motion. For the inner case $i=1$, while for the outer case $i=2$.  The index n refers to the numerical integration of the full equations of motion. \label{fig2}}
\end{figure}
\end{center}

\begin{center}
\begin{figure}
\includegraphics[width=90mm,height=50mm]{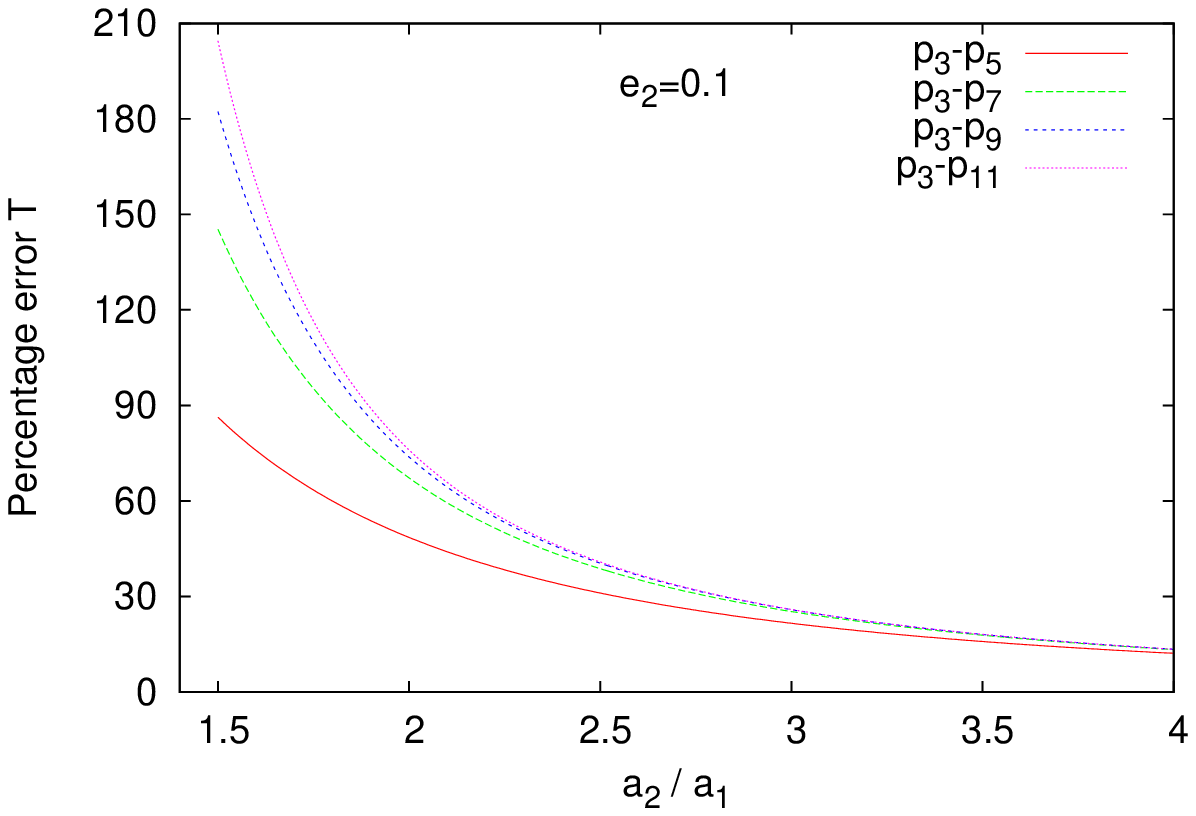}
\includegraphics[width=90mm,height=50mm]{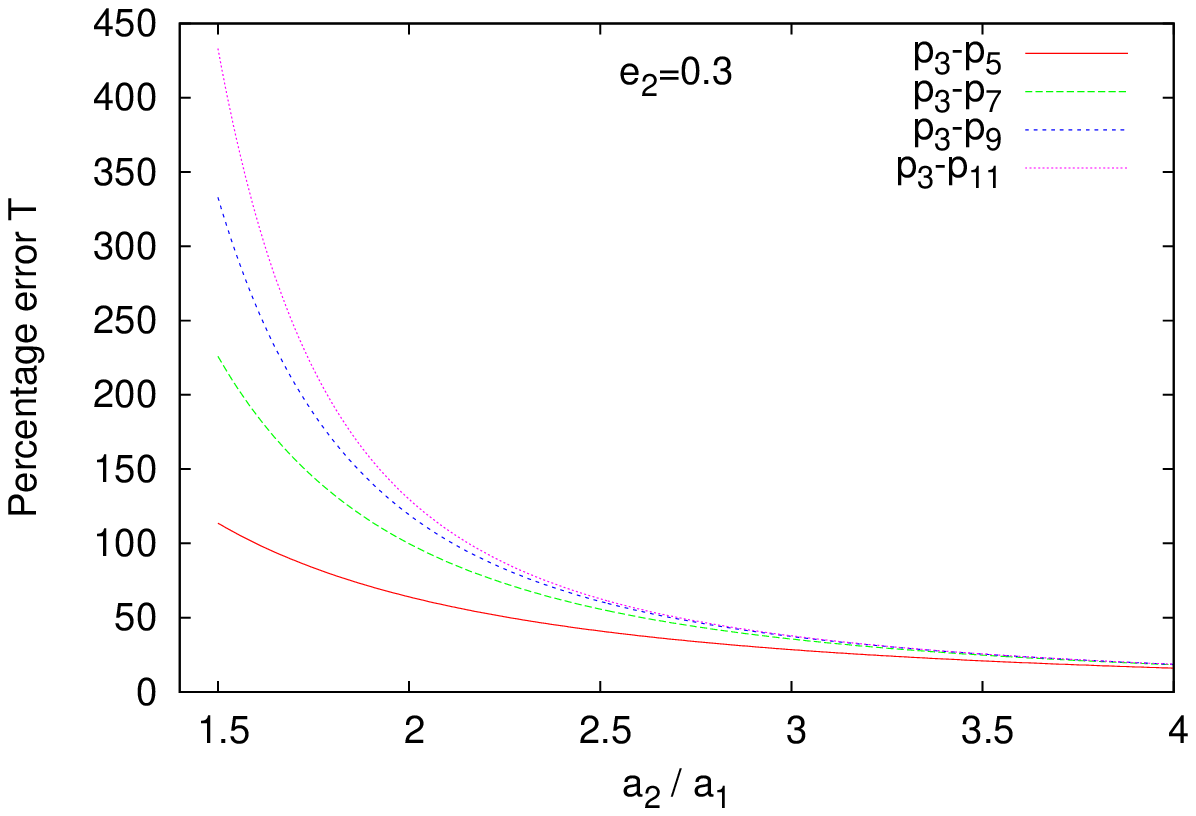}
\includegraphics[width=90mm,height=50mm]{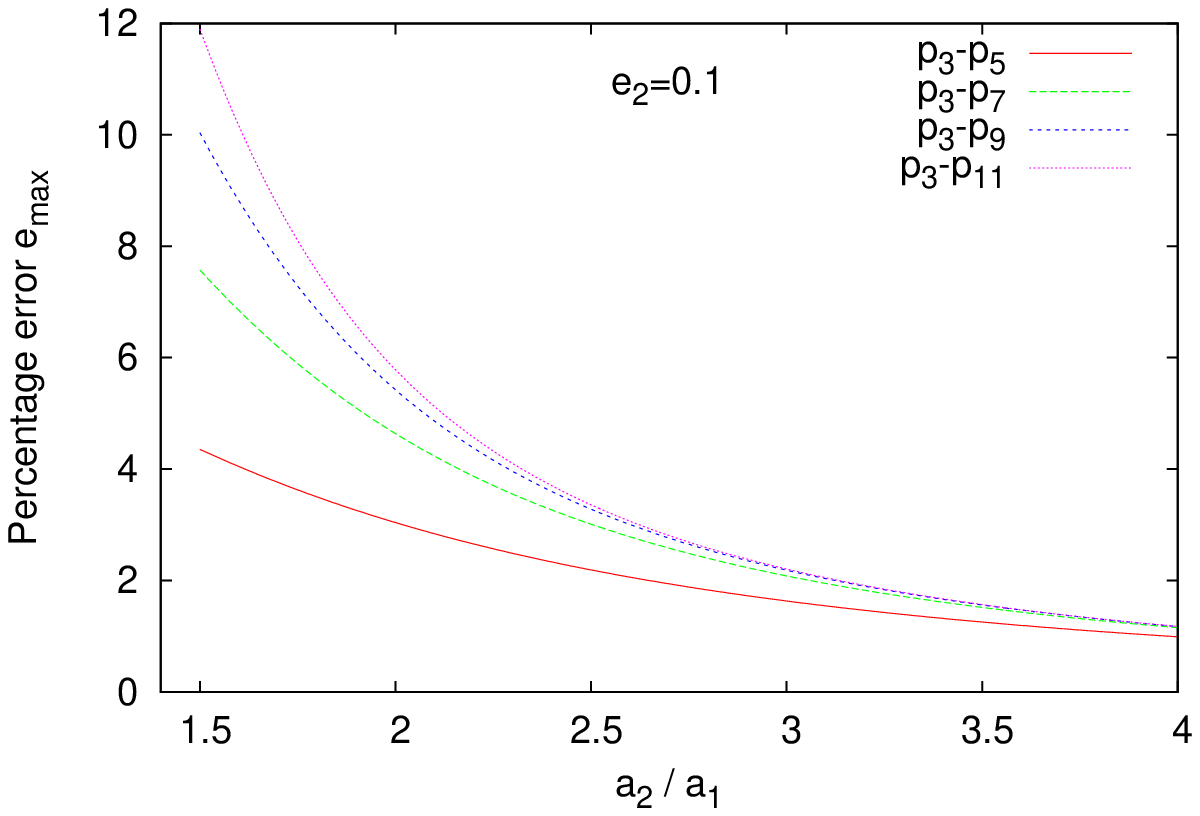}
\includegraphics[width=90mm,height=50mm]{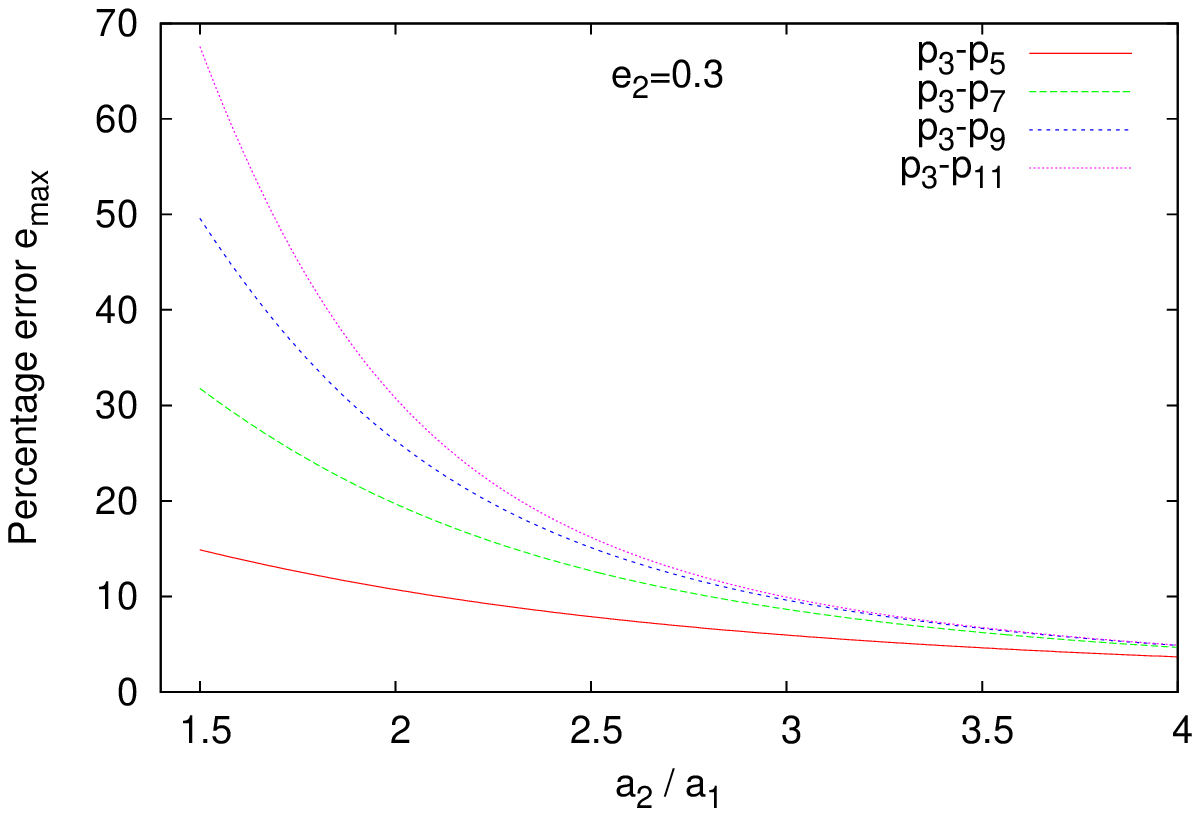}
\caption{Percentage error for the period T and maximum eccentricity $e_{max}$ between analytical solutions based on different expansion orders for a system with $m_0=1 M_{\odot}, m_1=1 M_{\oplus}, m_2=1 M_J$. For the left column plots we have $e_2=0.1$, while for the right ones we have $e_2=0.3$. The error curves shown here are the result of the comparison between the following analytical models : $P_3-P_5$, $P_3-P_7$, $P_3-P_9$, $P_3-P_{11}$, where ${P_n}$ $(n=3,5,7,9,11)$ denotes an analytical solution that is based on a Legendre polynomial expansion to order n.
\label{fig3}}
\end{figure}
\end{center}

\begin{center}
\begin{figure}
\includegraphics[width=90mm,height=50mm]{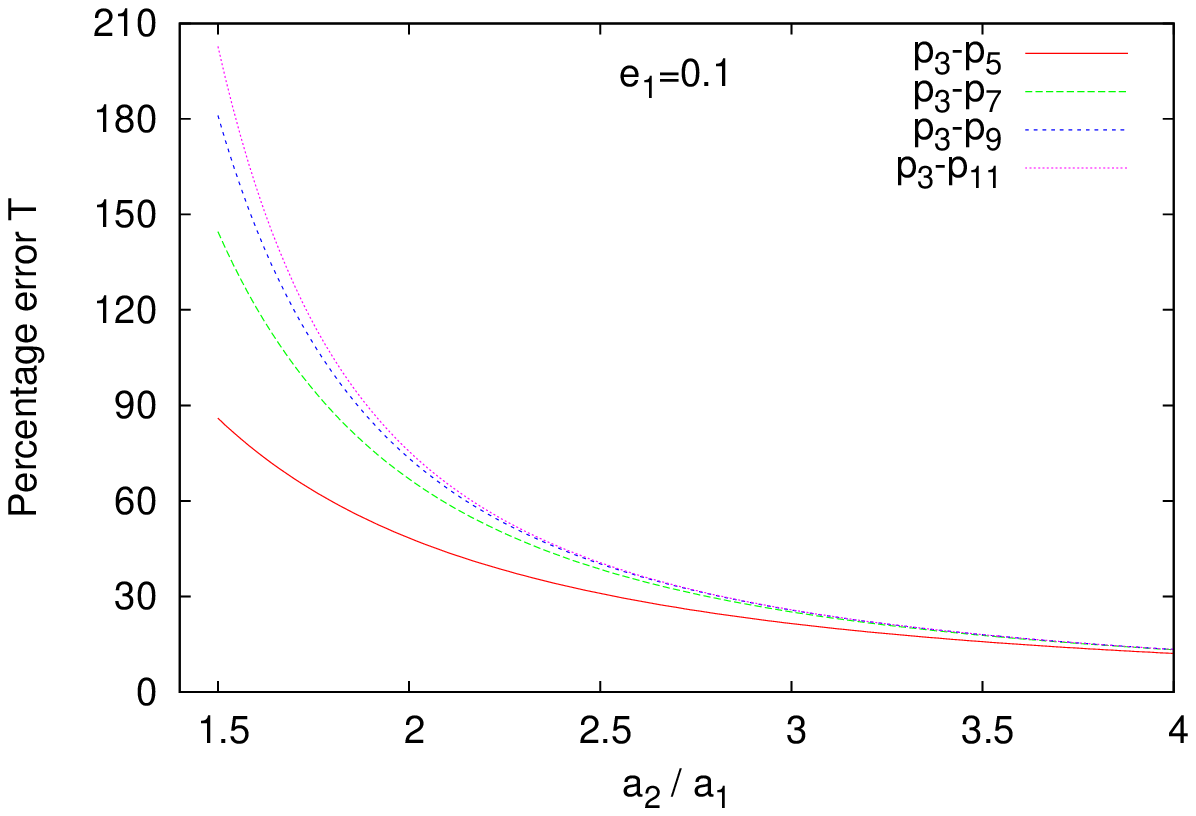}
\includegraphics[width=90mm,height=50mm]{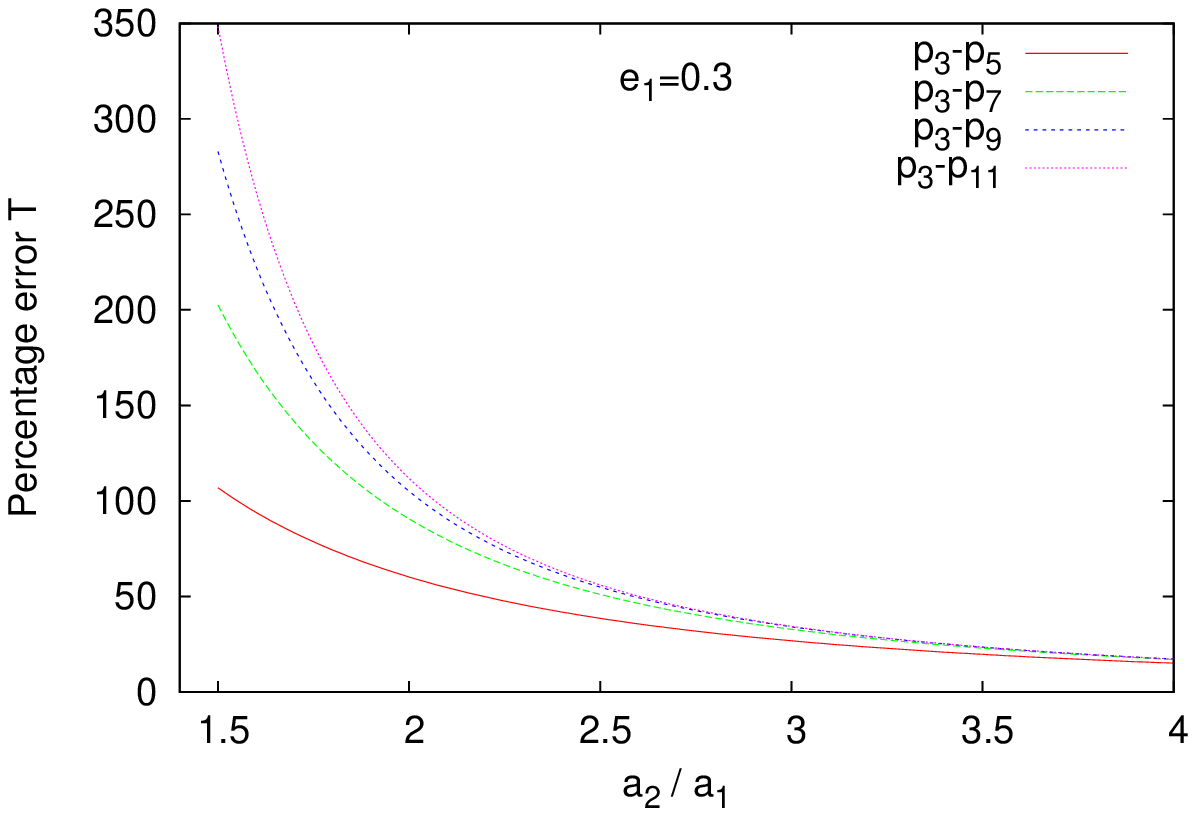}
\includegraphics[width=90mm,height=50mm]{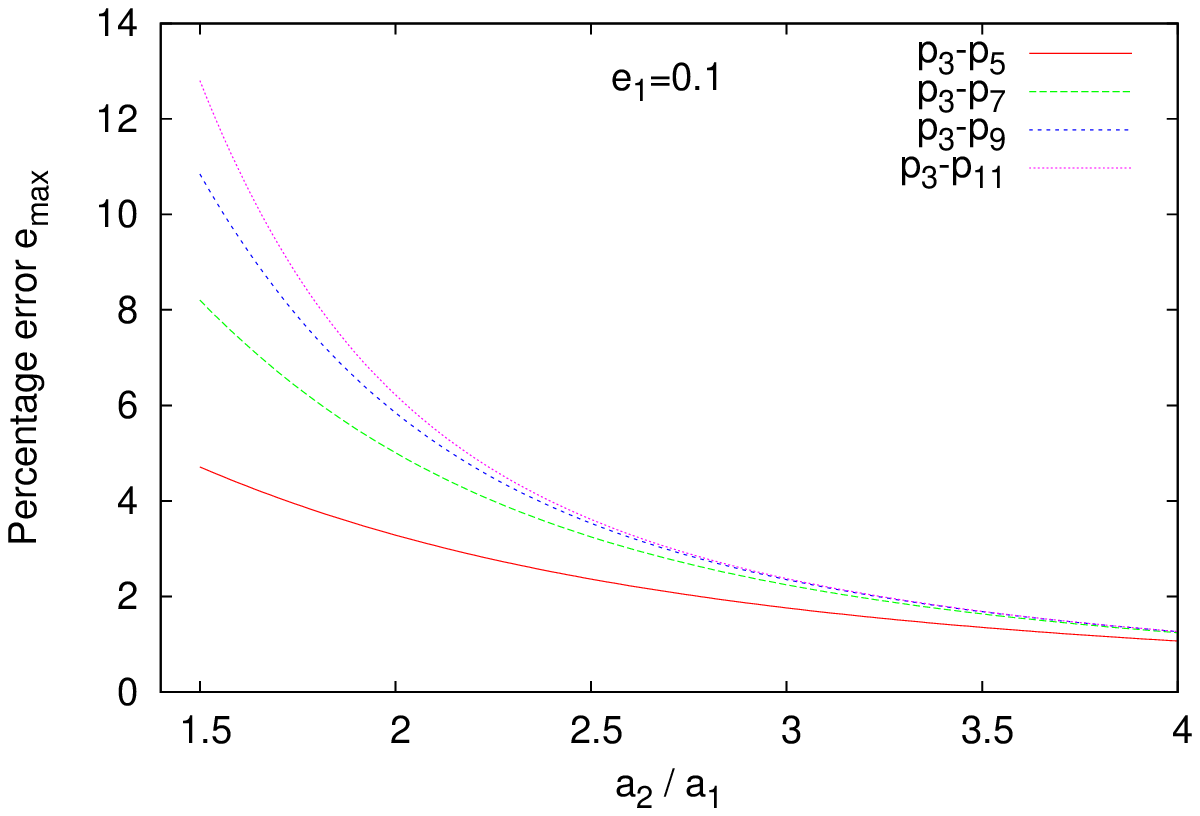}
\includegraphics[width=90mm,height=50mm]{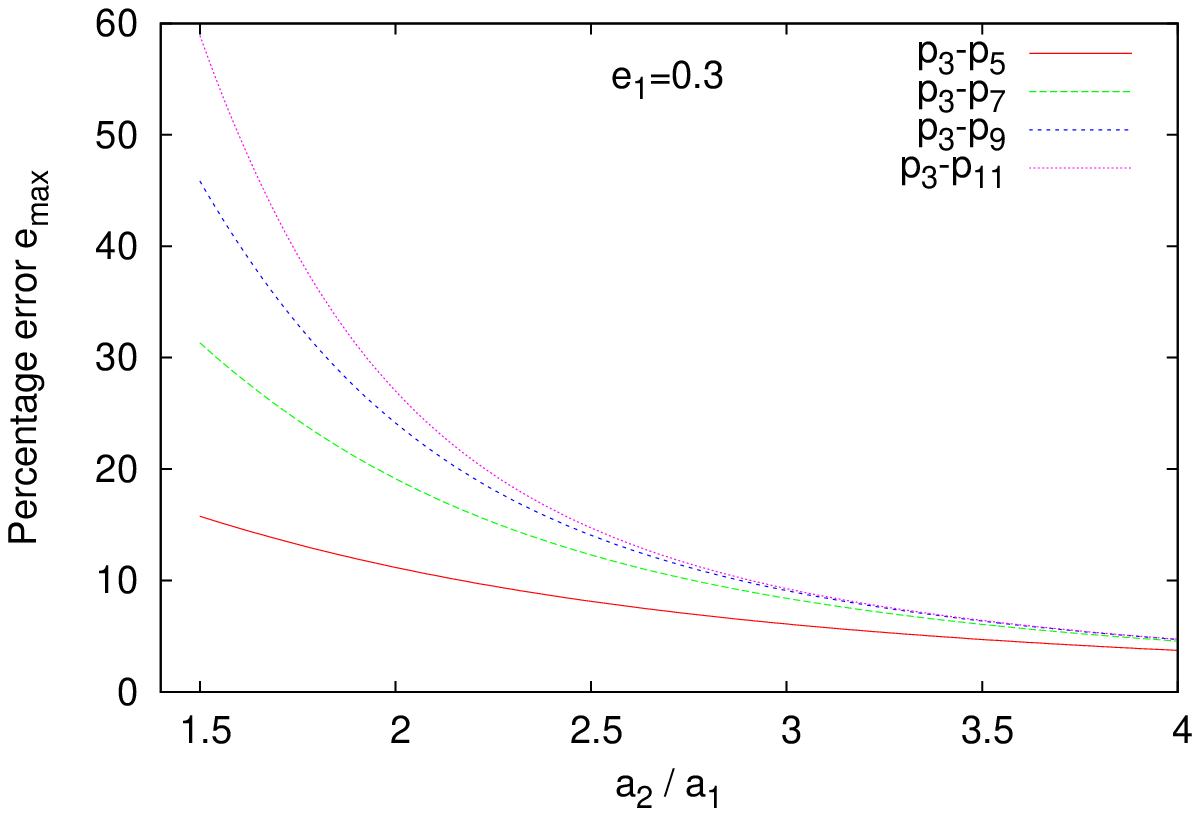}
\caption{Percentage error for the period T and maximum eccentricity $e_{max}$ between analytical solutions based on different expansion orders for a system with $m_0=1 M_{\odot}, m_1=1 M_J, m_2=1 M_{\oplus}$. For the left column plots we have $e_1=0.1$, while for the right ones we have $e_1=0.3$. The error curves shown here are the result of the comparison between the following analytical models : $P_3-P_5$, $P_3-P_7$, $P_3-P_9$, $P_3-P_{11}$, where ${P_n}$ $(n=3,5,7,9,11)$ denotes an analytical solution that is based on a Legendre polynomial expansion to order n. \label{fig4}}
\end{figure}
\end{center}

Figures \ref{fig2}, \ref{fig3} and \ref{fig4} present some of our findings for both inner and outer cases for a perturber of one Jupiter mass.  Figure 2 shows the logarithmic relative errors in the averaged square and maximum eccentricity, accompanied by a plot that shows the achieved maximum eccentricity.  The left column refers to the inner
case, while the right one refers to the outer case.  The first row shows the relative error between the analytical estimate and
the averaged square eccentricity obtained from the numerical integration of the full equations of motion.  One should keep in mind that a relative error in $<e^2_{i}>$ could be larger by a factor of two compared to an error in the averaged eccentricity.  In the left side of the plots we have unstable orbits (escape or ejection of the terrestrial planet) denoted by the lack of a symbol or orbits for which the maximum perturbed eccentricity acquires high values. However, as we move to the right of the plots, the semi-major axis ratio becomes smaller and our analytic results improve significantly. 

When the maximum perturbed eccentricity gets larger than 0.2-0.25, the assumption that the eccentricity of the terrestrial planet remains small starts to fail, as the third order terrestrial planet eccentricity terms neglected in the secular Hamiltonian become important to the dynamical evolution of the system.  Hence, the analytical theory is expected to start failing in such cases.  However, sometimes the analytical results may show good agreement with the full model for systems for which the terrestrial eccentricity becomes greater than $\sim$0.25.  Unfortunately, that good agreement could be coincidental, as the omission of the third order terms in the Hamiltonian was compensated by some other effect (e.g. being close to a mean motion resonance).  In order to avoid drawing the wrong conclusion about the accuracy of
the analytical estimates, we created the second row plots, which present the logarithmic relative error in the averaged square eccentricity between the analytical estimates and the results from the numerical integration of the full eleventh order secular equations of motion.  Usually, when the eccentricity becomes greater than $\sim$0.25, the higher order eccentricity terms omitted in the approximate Hamiltonian lead to an analytical solution that overestimates the period and amplitude of the secular orbital elements.  Therefore only when the error between the analytical estimates and the results from the full equations and the error between the analytical estimates and the results from the full secular equations are both small we can accept that the theoretical estimates perform well. The gap in the upper left corner of those plots is because the perturbed eccentricity grew larger than one and the quantity $\sqrt{(1-e_i^2)}$, which appears in some denominators in the secular equations of motion, could not be defined in the set of real numbers and consequently the simulation stopped.

The third row plots compare the analytical estimate for the maximum eccentricity with the maximum eccentricity achieved during the numerical integration of the full equations of motion.  Again, the error is high for systems at the upper left area of the plots and it progressively gets smaller as we move to the right.  The fourth row of plots presents the relative logarithmic error between the analytical estimate for the maximum eccentricity and the maximum eccentricity obtained during the numerical integration of the full eleventh order secular equations of motion.  The reasons we included that row were the same as in the case of the averaged square eccentricity.  The last row shows the maximum numerical eccentricity we got when we integrated the full equations of motion.  Using these plots to identify regions of high eccentricity is also a useful indicator of where we may have problems with our analytical estimates.   

As we stated earlier, when the initial conditions of a system are in the vicinity of a mean motion 
resonance, then our theory fails.  As we move away from that area, then the results improve again.  In the first row plots in Figure 2,
we have indicated a few mean motion resonance locations that coincide with the sampling of the initial semi-major axis.  The resonant effect is demonstrated by a vertical line that is redder (an indication of the sudden change in the error) compared to the neighbouring area.  As we move to smaller semi-major axis ratios, it gets more difficult to avoid the resonances as they become stronger and cover larger areas of the parameter space.  That, along with the fact that the convergence of our Legendre polynomial series expansion gets much slower for smaller
semi-major axis ratios, makes things rather difficult for our estimates when $a_2/a_1\lesssim 1.9$. 

Figures \ref{fig3} and \ref{fig4}  show the improvement of the analytical estimates for both inner and outer cases as we add more terms to the expansion. They show the percentage error in the period and in the maximum eccentricity between analytical solutions that are estimated to different orders of the semi-major axis ratio.
More specifically, we calculated the quantity $100(K_3-K_i)/K_i$ with $i=5,7,9,11$, where K is either the period of the evolution of the eccentricity/longitude of pericentre, given by equation (91), or the maximum eccentricity and the indices denote the order of the expansion we have used.  The increase in the order of the expansion between an analytical solution and its first improvement is always two because the A and B terms of the analytical solution come from Legendre polynomials with even indices, while the C term only contains odd order Legendre polynomials.  The reader should keep in mind that changes in the A term affect only the period, changes in the B term affect both the period and the maximum value of the eccentricity and changes in the C term affect only the maximum eccentricity. Therefore it is imperative that we move up in the expansion by two orders at a time in order to get the right picture. 

What we find by inspecting figures \ref{fig3} and \ref{fig4} is that the high order expansion is important for a more accurate picture of the evolution of the system as the semi-major axis ratio $a_2/a_1$ decreases.  The effect of the inclusion of higher order terms in the analytical solution is significantly more evident in the period than in the maximum value of the eccentricity.  In addition, as the eccentricity of the perturber increases, so does the effect of the inclusion of higher order terms.  In all cases, an expansion up to the third order is insufficient for the dynamical description of the system and the use of an expansion at least up to the fifth order is necessary, even for systems with larger semi-major axis ratios.  Finally, we would like to point out that the plots in figures \ref{fig3} and \ref{fig4} are not affected by a change in the mass of the perturber.  This is to be expected as the equations derived in the previous sections suggest.
  
More results from our simulations can be found in appendix \ref{A2}, where the reader may find similar plots as in figure 2, but for a perturbing body of 10, 0.1 and 0.03 Jupiter masses.  The results for those other masses show analogous behaviour as the results for a system where the perturber had a mass of $1 M_J$.  The main difference is that as the mass of the 
perturber gets smaller (larger) and hence the perturbation gets weaker (stronger), our analytical estimates are a good approximation to the full problem over a larger (smaller) area of our parameter space.

\section{Summary and discussion}
In this work we obtained an analytical solution for the long term motion of a smaller planet under the gravitational perturbations of a larger one with all bodies lying in the same plane.  The smaller planet was on a circular orbit initially.  However, the theory developed in this paper can apply to any system for which the terrestrial planet has a small non-zero initial eccentricity.  The different initial conditions are introduced into our equations through the constants $C_1$ and $C_2$.
 
We performed a high order expansion of the perturbing Hamiltonian in terms of the semi-major axis ratio of the two planets and we solved the derived equations of motion analytically.  The expansion and the analytical solution to the equations of motion have been given in a way that, depending on the application and the required accuracy, the desired order in the semi-major axis ratio and the eccentricity of the perturber can be retained.  Moreover, the new secular solution can be easily combined, when required, with short period terms \citep[as for example in][]{Georgakarakos-2003}.  Such terms in the types of systems we investigate become necessary when the eccentricity of the perturber is low ($\lesssim0.01$).  The analytical solution was then compared with results obtained from numerical simulations and the agreement between the analytical and the numerical results was generally good.  

As expected, the new analytical results  do not work well near mean motion resonances, 
when the two planetary bodies have comparable masses and consequently the eccentricity and pericentre of the perturber can not be assumed constant any longer, and in cases where the pericentre of the larger body changes over time due to general relativity.
Such cases must be treated in a different manner.  For instance, in the case where the giant planet orbit is inside that of the smaller planet it may be required to include a post Newtonian correction to the motion of the pericentre of the giant planet in order to model the system more accurately.  In such a scenario, equations (\ref{de3}) and (\ref{de4}) will no longer have constant coefficients and a different type of solution should be sought.  On the other hand, though planets near a mean motion resonance could completely ruin our solution as we move away from the location of the mean motion resonance, the secular solution may provide a decent approximation to the motion of the system, but it may have to be coupled with some mean motion resonance modelling to improve its accuracy.
Finally, the situation will get even more complicated if we assume that the longitude of pericentre of the perturber is
affected by the perturbed body.  This is a rather necessary assumption to be made when dealing with two planets of comparable size, and some test simulations we ran confirmed this.
In this situation, contrary to what happens with the inclusion of the post-Newtonian correction which normally produces a constant rate of change for the pericentre motion, the rate of change of the pericentre will be a non-constant function of time and finding the solution for those differential equations could be hard.

As mentioned earlier, the number of discovered exoplanets has recently risen above three thousand. Detection methods have been improved sufficiently over the years that we are
currently able to detect planets of comparable size to that of Earth, though much closer to the host star \citep[e.g.][]{2014ApJS..210...20M}.  
The scientific community has become more interested than ever to answer the question of whether life as
we know it can flourish elsewhere in the universe and as a result, a large number of papers discussing habitability have emerged.  Planetary eccentricity is an important aspect of habitability, as it can affect the climate of the planet
\cite[e.g.][]{2002IJAsB...1...61W,2010ApJ...721.1308S,2010ApJ...721.1295D,2015P&SS..105...43L,eggl-et-al-2012}.  Therefore,
the new analytical estimates presented in the previous sections could be used in such studies.
Our future plans include improving the solution for the long term motion by taking into consideration other dynamical effects, such as mean motion resonances or general relativity and trying to provide orbital solutions for systems with comparable planetary masses.

\section*{Acknowledgements}

We would like to thank the anonymous referee for his/her useful comments that helped
us improve the manuscript. We would also like to thank the High Performance Computing Resources team at New York University Abu Dhabi and especially Jorge Naranjo for helping us with our numerical simulations.


 

\bibliographystyle{mnras}
\bibliography{ref}


\appendix
\section{Equations of motion}
\label{A1}
For simplicity, we have dropped all the indices.\\

\noindent Since $G=L\sqrt{1-e^2}$, we have:
\begin{eqnarray}
\noindent\frac{dG}{dt}&=&-\frac{\partial H}{dg}\Rightarrow\sqrt{1-e^2}\frac{dL}{dt}-\frac{Le}{\sqrt{1-e^2}}\frac{de}{dt}=-\frac{\partial H}{dg}\Rightarrow\frac{de}{dt}=\frac{\sqrt{1-e^2}}{Le}(\sqrt{1-e^2}\frac{dL}{dt}+\frac{\partial H}{dg})\Rightarrow\nonumber\\
\Rightarrow\frac{de}{dt}&=&\frac{\sqrt{1-e^2}}{Le}(-\sqrt{1-e^2}\frac{\partial H}{\partial l}+\frac{\partial H}{dg}).
\end{eqnarray}
Since the Hamiltonian is independent of $l$, then:
\begin{eqnarray}
\frac{de}{dt}=\frac{\sqrt{1-e^2}}{Le}\frac{\partial H}{dg}.
\end{eqnarray} 
\begin{eqnarray}
\frac{d\varpi}{dt}&=&\frac{\partial H}{\partial G}\Rightarrow\frac{d\varpi}{dt}=\frac{\partial H}{\partial e}\frac{\partial e}{\partial G}\Rightarrow\frac{d\varpi}{dt}=-\frac{\sqrt{1-e^2}}{Le}\frac{\partial H}{\partial e}
\end{eqnarray}
\section{Complete eleventh order perturbing Hamiltonian}
\label{ham11}
\begin{eqnarray}
\noindent H_{p_2} &=& -\frac{1}{4}\frac{\mathcal{G}m_0m_1m_2}{(m_0+m_1)(1-e_{2}^2)^{3/2}}\frac{a_{1}^2}{a_{2}^3}(1+\frac{3}{2}e_{1}^2)\\
H_{p_3} &=& \frac{15}{16}\frac{\mathcal{G}m_0m_1m_2(m_0-m_1)}{(m_0+m_1)^2(1-e_{2}^2)^{5/2}}\frac{a_{1}^3}{a_{2}^4}e_{1}e_{2}(1+\frac{3}{4}e_{1}^2)\cos{(\varpi_{1}-\varpi_{2})}\\
H_{p_{4}} &=& -\frac{9}{64}\frac{\mathcal{G}m_0m_1m_2(m_0^3+m_1^3)}{(m_0+m_1)^4(1-e_{2}^2)^{7/2}}\frac{a_{1}^4}{a_{2}^5}\bigg[1+5e_{1}^2+\frac{15}{8}e_{1}^4+(\frac{3}{2}+\frac{15}{2}e_{1}^2+\frac{45}{16}e_{1}^4)e_{2}^2+(\frac{35}{4}+\frac{35}{8}e_{1}^2)e_{1}^2e_{2}^2\times\nonumber\\
& & \times\cos{(2\varpi_{1}-2\varpi_{2})}\bigg]\\
H_{p_{5}} &=& \frac{\mathcal{G}m_0m_1m_2(m_0^4-m_1^4)}{(m_0+m_1)^5(1-e_{2}^2)^{9/2}}\frac{a_{1}^5}{a_{2}^6}\bigg[\big[(\frac{315}{256}e_{1}+\frac{1575}{2048}e_{1}^5+\frac{1575}{512}e_{1}^3)e_{2}^3+(\frac{105}{64}e_{1}+\frac{525}{512}e_{1}^5+\frac{525}{128}e_{1}^3)e_{2}\big]\cos(\varpi_{1}-\varpi_{2})+\nonumber\\
& & +(\frac{735}{512}e_{1}^3+\frac{2205}{4096}e_{1}^5)e_{2}^3\cos(3\varpi_{1}-3\varpi_{2})\bigg]\\
H_{p_{6}} &=&-\frac{\mathcal{G}m_0m_1m_2(m_0^5+m_1^5)}{(m_0+m_1)^6(1-e_{2}^2)^{11/2}}\frac{a_{1}^6}{a_{2}^7}\bigg[\big[(-\frac{23625}{32768}e_{1}^6-\frac{7875}{2048}e_{1}^4-\frac{4725}{2048}e_{1}^2)e_{2}^4+(-\frac{7875}{1024}e_{1}^4-\frac{23625}{16384}e_{1}^6-\frac{4725}{1024}e_{1}^2)e_{2}^2\big]\times\nonumber\\
& & \times\cos(2\varpi_{1}-2\varpi_{2})+(-\frac{51975}{32768}e_{1}^4-\frac{31185}{65536}e_{1}^6)e_{2}^4\cos(4\varpi_{1}-4\varpi_{2})+(-\frac{375}{2048}-\frac{39375}{16384}e_{1}^4-\frac{13125}{32768}e_{1}^6-\frac{7875}{4096}e_{1}^2)e_{2}^4-\nonumber\\
& &-\frac{2625}{2048}e_{1}^4-\frac{875}{4096}e_{1}^6+(-\frac{4375}{4096}e_{1}^6-\frac{125}{256}-\frac{13125}{2048}e_{1}^4-\frac{2625}{512}e_{1}^2)e_{2}^2
-\frac{25}{256}-\frac{525}{512}e_{1}^2\bigg]\\
H_{p_{7}} &=&\frac{\mathcal{G}m_0m_1m_2(m_0^6-m_1^6)}{(m_0+m_1)^7(1-e_{2}^2)^{13/2}}\frac{a_{1}^7}{a_{2}^8}\bigg[\big[(\frac{23625}{16384}e_{1}+\frac{826875}{1048576}e_{1}^7+\frac{826875}{131072}e_{1}^5+\frac{496125}{65536}e_{1}^3)e_{2}^5+(\frac{826875}{32768}e_{1}^5+\frac{496125}{16384}e_{1}^3+\nonumber\\
& & +\frac{23625}{4096}e_{1}+\frac{826875}{262144}e_{1}^7)e_{2}^3+(
\frac{165375}{131072}e_{1}^7+\frac{165375}{16384}e_{1}^5+\frac{99225}{8192}e_{1}^3+\frac{4725}{2048}e_{1})e_{2}\big]\cos(\varpi_{1}-\varpi_{2})+\big[(\frac{1403325}{2097152}e_{1}^7+\nonumber\\
& & +\frac{2338875}{524288}e_{1}^5+\frac{467775}{131072}e_{1}^3)e_{2}^5+(\frac{155925}{16384}e_{1}^3+\frac{779625}{65536}e_{1}^5+\frac{467775}{262144}e_{1}^7)e_{2}^3\big]\cos(3\varpi_{1}-3\varpi_{2})+(\frac{891891}{2097152}e_{1}^7+\nonumber\\
& & +\frac{891891}{524288}e_{1}^5)e_{2}^5\cos(5\varpi_{1}-5\varpi_{2})\bigg]\\
H_{p_{8}} &=& -\frac{\mathcal{G}m_0m_1m_2(m_0^7+m_1^7)}{(m_0+m_1)^8(1-e_{2}^2)^{15/2}}\frac{a_{1}^8}{a_{2}^9}\bigg[\big[(\frac{12733875}{16777216}e_{1}^8+\frac{63669375}{8388608}e_{1}^6+\frac{12733875}{1048576}e_{1}^4+\frac{1819125}{524288}e_{1}^2)e_{2}^6+(\frac{4244625}{1048576}e_{1}^8+\nonumber\\
& & +\frac{606375}{32768}e_{1}^2+\frac{21223125}{524288}e_{1}^6
+\frac{4244625}{65536}e_{1}^4)e_{2}^4+(\frac{2546775}{1048576}e_{1}^8+\frac{2546775}{65536}e_{1}^4+\frac{12733875}{524288}e_{1}^6+\frac{363825}{32768}e_{1}^2)e_{2}^2\big]\cos{(2\varpi_{1}-2\varpi_{2})}+\nonumber\\
& & +\big[(\frac{17342325}{1048576}e_{1}^6+\frac{17342325}{8388608}e_{1}^8+\frac{17342325}{1048576}e_{1}^4)e_{2}^4+(\frac{10405395}{2097152}e_{1}^6+\frac{10405395}{2097152}e_{1}^4+\frac{10405395}{16777216}e_{1}^8)e_{2}^6\big]\cos(4\varpi_{1}-4\varpi_{2})+\nonumber\\
& & +(\frac{15030015}{8388608}e_{1}^6+\frac{6441435}{16777216}e_{1}^8)e_{2}^6\cos(6\varpi_{1}-6\varpi_{2})+(\frac{385875}{131072}e_{1}^2+\frac{13505625}{33554432}e_{1}^8+\frac{4501875}{1048576}e_{1}^6+\frac{42875}{262144}+\frac{8103375}{1048576}e_{1}^4)e_{2}^6+\nonumber\\
& & +\frac{128625}{65536}e_{1}^6+\frac{385875}{2097152}e_{1}^8+\frac{11025}{8192}e_{1}^2+(\frac{13505625}{524288}e_{1}^6+\frac{1157625}{65536}e_{1}^2+\frac{40516875}{16777216}e_{1}^8+\frac{128625}{131072}+
\frac{24310125}{524288}e_{1}^4)e_{2}^4+\frac{1225}{16384}+\nonumber\\
& & +\frac{25725}{32768}+\frac{8103375}{4194304}e_{1}^8+\frac{4862025}{131072}e_{1}^4+\frac{2701125}{131072}
e_{1}^6+\frac{231525}{16384}e_{1}^2)e_{2}^2+\frac{231525}{65536}e_{1}^4\bigg]\\
H_{p_{9}} &=& \frac{\mathcal{G}m_0m_1m_2(m_0^8-m_1^8)}{(m_0+m_1)^9(1-e_{2}^2)^{17/2}}\bigg[\big[(\frac{53482275}{67108864}e_{1}^9+\frac{53482275}{2097152}e_{1}^5+\frac{7640325}{524288}e_{1}^3+\frac{89137125}{8388608}e_{1}^7+\frac{848925}{524288}e_{1})e_{2}^7+(\frac{53482275}{262144}e_{1}^5+\nonumber\\
& & +\frac{89137125}{1048576}e_{1}^7+\frac{848925}{65536}e_{1}+\frac{53482275}{8388608}e_{1}^9+\frac{7640325}{65536}e_{1}^3)e_{2}^5+(\frac{4584195}{32768}e_{1}^3+\frac{509355}{32768}e_{1}+\frac{32089365}{131072}e_{1}^5+\frac{32089365}{4194304}e_{1}^9+\nonumber\\
& & +\frac{53482275}{524288}e_{1}^7)e_{2}^3+(
\frac{1528065}{1048576}e_{1}^9+\frac{1528065}{32768}e_{1}^5+\frac{218295}{8192}e_{1}^3+\frac{2546775}{131072}e_{1}^7+\frac{24255}{8192}e_{1})e_{2}\big]\cos{(\varpi_{1}-\varpi_{2})}+\big[(\frac{24279255}{33554432}e_{1}^9+\nonumber\\
& & +\frac{72837765}{4194304}e_{1}^5+\frac{3468465}{524288}
e_{1}^3+\frac{72837765}{8388608}e_{1}^7)e_{2}^7+(\frac{121396275}{2097152}e_{1}^7+
\frac{40465425}{8388608}e_{1}^9+\frac{121396275}{1048576}e_{1}^5+\frac{5780775}{131072}
e_{1}^3)e_{2}^5+\nonumber\\
& & +(\frac{24279255}{524288}e_{1}^7+\frac{24279255}{262144}e_{1}^5+\frac{1156155}{32768}e_{1}^3+\frac{8093085}{2097152}e_{1}^9)e_{2}^3\big]\cos{3\varpi_{1}-3\varpi)}+\big[(\frac{19324305}{33554432}e_{1}^9+\frac{45090045}{8388608}e_{1}^7+\nonumber\\
& & +\frac{27054027}{4194304}e_{1}^5)e_{2}^7+(\frac{45090045}{2097152}e_{1}^7+\frac{19324305}{8388608}e_{1}^9+\frac{27054027}{1048576}e_{1}^5)e_{2}^5\big]\cos{(5\varpi_{1}-5\varpi_{2})}+(\frac{15643485}{8388608}e_{1}^7+\nonumber\\
& & +\frac{46930455}{134217728}e_{1}^9)e_{2}^7\cos{(7\varpi_{1}-7\varpi_{2})}\bigg]\\
H_{p_{10}} &=& -\frac{\mathcal{G}m_0m_1m_2(m_0^9+m_1^9)}{(m_0+m_1)^{10}(1-e_{2}^2)^{19/2}}\frac{a_{1}^{10}}{a_{2}^{11}}\bigg[\big[(\frac{297972675}{1048576}e_{1}^4+\frac{99324225}{2097152}e_{1}^2+\frac{2085808725}{16777216}e_{1}^8+\frac{2085808725}{268435456}e_{1}^{10}+\frac{6257426175}{16777216}e_{1}^6)e_{2}^6+\nonumber\\
& & +(\frac{417161745}{536870912}e_{1}^{10}+\frac{19864845}{4194304}e_{1}^2+\frac{59594535}{2097152}e_{1}^4+\frac{1251485235}{33554432}e_{1}^6+\frac{417161745}{33554432}e_{1}^8)e_{2}^8+(\frac{59594535}{16777216}e_{1}^{10}+\frac{8513505}{65536}e_{1}^4+\nonumber
\end{eqnarray}
\begin{eqnarray}
& & +\frac{2837835}{131072}e_{1}^2+\frac{59594535}{1048576}e_{1}^8+\frac{178783605}{1048576}e_{1}^6)e_{2}^2+(\frac{417161745}{33554432}e_{1}^{10}+\frac{1251485235}{2097152}e_{1}^6+\frac{59594535}{131072}e_{1}^4+\frac{19864845}{262144}e_{1}^2+\nonumber\\
& & +\frac{417161745}{2097152}e_{1}^8)e_{2}^4\big]\cos{(2\varpi_{1}-2\varpi_{2})}+\big[(\frac{1291214925}{134217728}e_{1}^8+\frac{184459275}{16777216}e_{1}^4+\frac{184459275}{268435456}e_{1}^{10}+\frac{774728955}{33554432}e_{1}^6)e_{2}^8+\nonumber\\
& & +(\frac{184459275}{2097152}e_{1}^4+\frac{184459275}{33554432}e_{1}^{10}+\frac{1291214925}{16777216}e_{1}^8+\frac{774728955}{4194304}e_{1}^6)e_{2}^4+(\frac{184459275}{2097152}e_{1}^4+\frac{184459275}{33554432}e_{1}^{10}+\nonumber\\
& & +\frac{1291214925}{16777216}e_{1}^8+\frac{774728955}{4194304}e_{1}^6)e_{2}^6\big]\cos{(4\varpi_{1}-4\varpi_{2})}+\big[(\frac{191988225}{33554432}e_{1}^8+\frac{575964675}{1073741824}e_{1}^{10}+\frac{268783515}{33554432}e_{1}^6)e_{2}^8+\nonumber\\
& & +(\frac{447972525}{16777216}e_{1}^8+\frac{627161535}{16777216}e_{1}^6+\frac{1343917575}{536870912}e_{1}^{10})e_{2}^6\big]\cos{(6\varpi_{1}-6\varpi_{2})}+(\frac{1378048815}{4294967296}e_{1}^{10}+\nonumber\\
& & +\frac{4134146445}{2147483648}e_{1}^8)e_{2}^8\cos{(8\varpi_{1}-8\varpi_{2})}+(\frac{7220107125}{1073741824}e_{1}^8+\frac{1444021425}{67108864}e_{1}^6+\frac{1250235}{8388608}+\frac{618866325}{33554432}e_{1}^4+\frac{68762925}{16777216}e_{1}^2+\nonumber\\
& & +\frac{866412855}{2147483648}e_{1}^{10})e_{2}^8+(\frac{22920975}{524288}e_{1}^2+\frac{288804285}{67108864}e_{1}^{10}+\frac{416745}{262144}+
\frac{206288775}{1048576}e_{1}^4+\frac{2406702375}{33554432}e_{1}^8+\frac{481340475}{2097152}e_{1}^6)e_{2}^6+\nonumber\\
& & +\frac{3969}{65536}+\frac{218295}{131072}e_{1}^2+
\frac{2750517}{16777216}e_{1}^{10}+\frac{4584195}{524288}e_{1}^6\frac{1964655}{262144}e_{1}^4+(\frac{17681895}{131072}e_{1}^4+\frac{24754653}{8388608}e_{1}^{10}+
\frac{41257755}{262144}e_{1}^6+\frac{1964655}{65536}e_{1}^2+\nonumber\\
& & +\frac{35721}{32768}+\frac{206288775}{4194304}e_{1}^8)e_{2}^2+(\frac{866412855}{2097152}e_{1}^6+
\frac{519847713}{67108864}e_{1}^{10}+\frac{4332064275}{33554432}e_{1}^8+\frac{371319795}{1048576}e_{1}^4+\frac{750141}{262144}+\nonumber\\
& & \frac{41257755}{524288}e_{1}^2)e_{2}^4+\frac{22920975}{8388608}e_{1}^8\bigg]\\
H_{p_{11}} &=& \frac{\mathcal{G}m_0m_1m_2(m_0^{10}-m_1^{10})}{(m_0+m_1)^{11}(1-e_{2}^2)^{21/2}}\frac{a_{1}^{11}}{a_{2}^{12}}\bigg[\big[
(\frac{68831687925}{4294967296}e_{1}^9+\frac{9833098275}{134217728}e_{1}^5+\frac{68831687925}{1073741824}e_{1}^7+\frac{13766337585}{17179869184}e_{1}^{11}+\frac{3277699425}{134217728}e_{1}^3+\nonumber\\
& & +\frac{59594535}{33554432}e_{1})e_{2}^9+(\frac{22943895975}{2147483648}e_{1}^{11}+\frac{99324225}{4194304}e_{1}+\frac{5462832375}{16777216}e_{1}^3+\frac{16388497125}{16777216}e_{1}^5+\frac{114719479875}{536870912}e_{1}^9+\nonumber\\
& & +\frac{114719479875}{134217728}e_{1}^7)e_{2}^7+(\frac{9833098275}{4194304}e_{1}^5+\frac{68831687925}{134217728}e_{1}^9+\frac{3277699425}{4194304}e_{1}^3+\frac{68831687925}{33554432}e_{1}^7+\frac{13766337585}{536870912}e_{1}^{11}+\nonumber\\
& & +\frac{59594535}{1048576}e_{1})e_{2}^5+(\frac{8513505}{262144}e_{1}+\frac{468242775}{1048576}e_{1}^3+\frac{9833098275}{33554432}e_{1}^9+\frac{9833098275}{8388608}e_{1}^7+\frac{1966619655}{134217728}e_{1}^{11}+\frac{1404728325}{1048576}e_{1}^5)e_{2}^3+\nonumber\\
& & +\frac{1}{68719476736}(8950304563200e_{1}^7+247973806080e_{1}+111878807040e_{1}^{11}+10228919500800e_{1}^5+2237576140800e_{1}^9+\nonumber\\
& & +3409639833600e_{1}^3)e2\big]\cos{(\varpi_{1}-\varpi_{2})}+\big[(\frac{6456074625}{8589934592}e_{1}^{11}+\frac{717341625}{67108864}e_{1}^3+\frac{6456074625}{134217728}e_{1}^5+\frac{15064174125}{1073741824}e_{1}^9+\nonumber\\
& & +\frac{27115513425}{536870912}e_{1}^7)e_{2}^9+(\frac{19368223875}{33554432}e_{1}^5+\frac{19368223875}{2147483648}e_{1}^{11}+\frac{81346540275}{134217728}e_{1}^7+\frac{45192522375}{268435456}e_{1}^9+\frac{2152024875}{16777216}e_{1}^3)e_{2}^7+\nonumber\\
& & +(\frac{19368223875}{16777216}e_{1}^5+\frac{19368223875}{1073741824}e_{1}^{11}+\frac{45192522375}{134217728}e_{1}^9+\frac{81346540275}{67108864}e_{1}^7\frac{2152024875}{8388608}e_{1}^3)e_{2}^5+\nonumber\\
& & +\frac{1}{68719476736}(6715957248000e_{1}^3+30221807616000e_{1}^5+8814693888000e_{1}^9+31732897996800e_{1}^7+\nonumber\\
& & +472215744000e_{1}^{11})e_{2}^3\big])\cos{(3\varpi_{1}-3\varpi_{2})}+\big[(\frac{15679038375}{536870912}e_{1}^7\frac{2239862625}{134217728}e_{1}^5+\frac{11199313125}{17179869184}e_{1}^{11}+\frac{11199313125}{1073741824}e_{1}^9)e_{2}^9+\nonumber\\
& & +(\frac{5226346125}{33554432}e_{1}^5+\frac{36584422875}{134217728}e_{1}^7+\frac{26131730625}{268435456}e_{1}^9+\frac{26131730625}{4294967296}e_{1}^{11})e_{2}^7+\frac{1}{68719476736}(22477469414400e_{1}^7+\nonumber\\
& & +8027667648000e_{1}^9+501729228000e_{1}^{11}+12844268236800e_{1}^5)e_{2}^5\big])\cos{(5\varpi_{1}-5\varpi_{2})}+\big[(\frac{103353661125}{17179869184}e_{1}^9+\frac{34451220375}{68719476736}e_{1}^{11}+\nonumber\\
& & +\frac{20670732225}{2147483648}e_{1}^7)e_{2}^9+\frac{1}{68719476736}(3527804966400e_{1}^7+2204878104000e_{1}^9+183739842000e_{1}^{11})e_{2}^7\big]\cos{(7\varpi_{1}-7\varpi_{2})}+\nonumber\\
& & +\frac{1}{68719476736}(135763327700e_{1}^9+20364499155e_{1}^{11})\cos{(9\varpi_{1}-9\varpi_{2})}e_{2}^9\bigg]
\end{eqnarray}

\section{Numerical results}
\label{A2}

\begin{center}
\begin{figure}
\includegraphics[width=92mm,height=120mm]{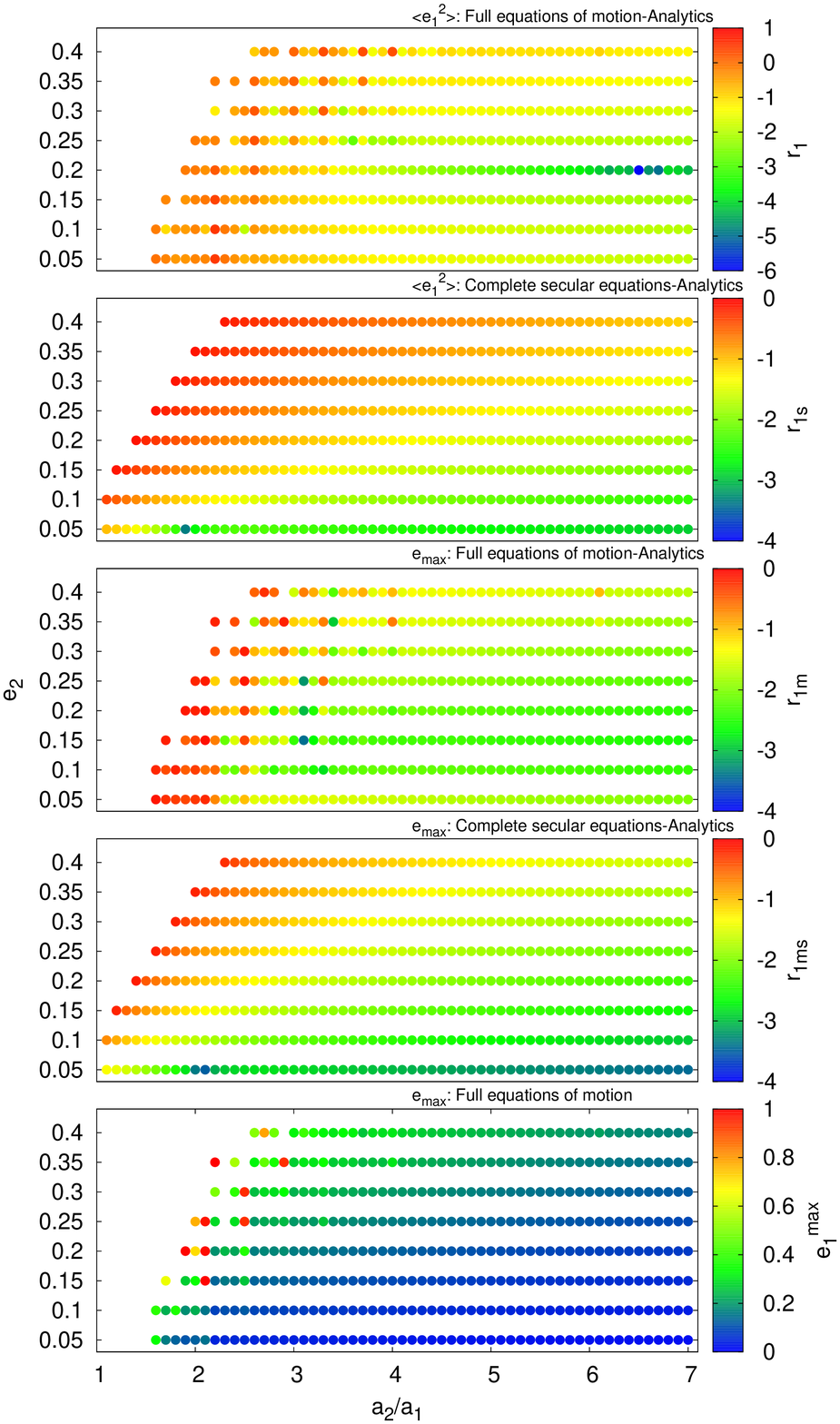}
\includegraphics[width=92mm,height=120mm]{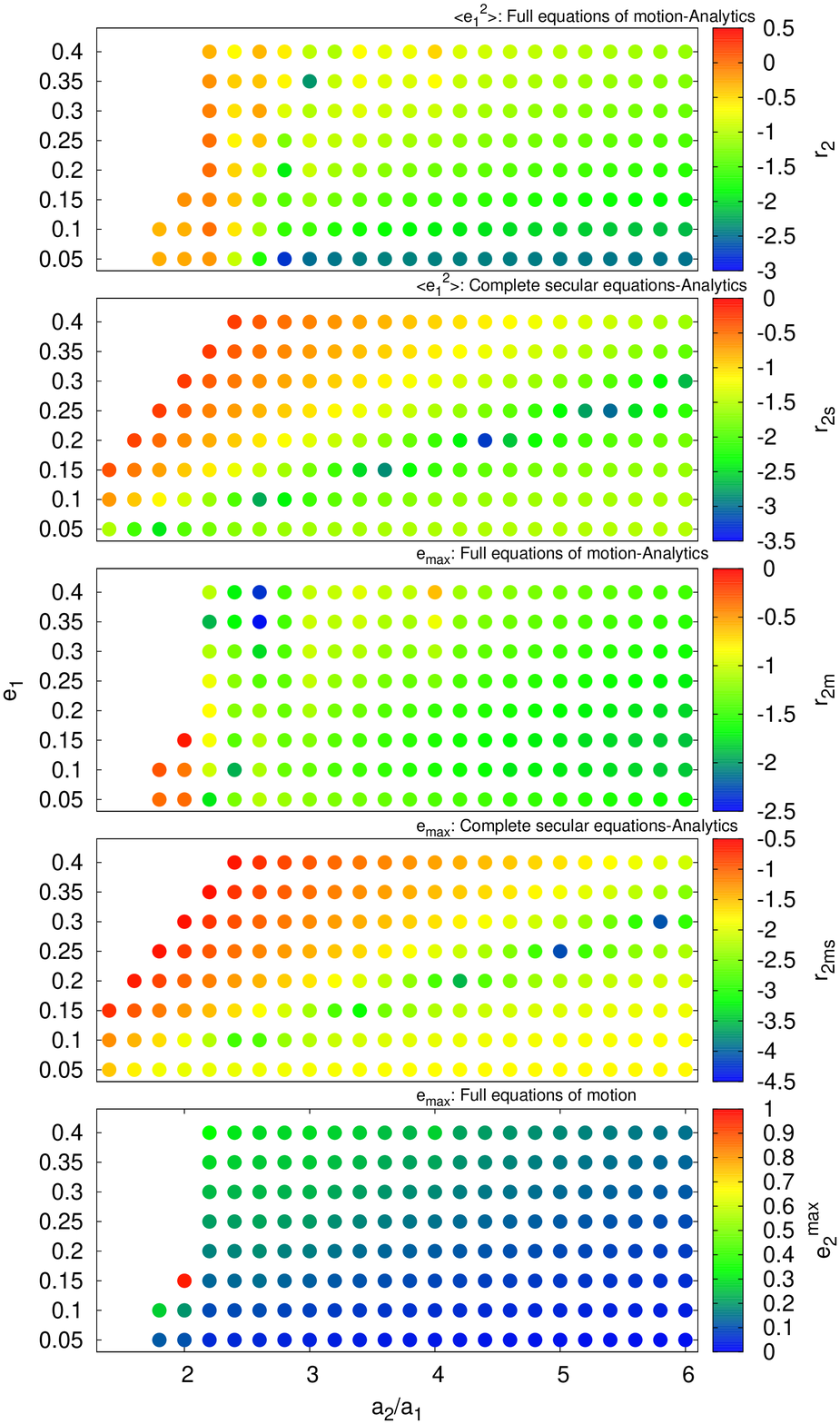}
\caption{Same as in figure 2, but for a system with $m_2=10 M_J$ for the inner case (left column)
and $m_1=10 M_J$ for the outer case (right column).}
\end{figure}
\end{center}

\begin{center}
\begin{figure}
\includegraphics[width=92mm,height=120mm]{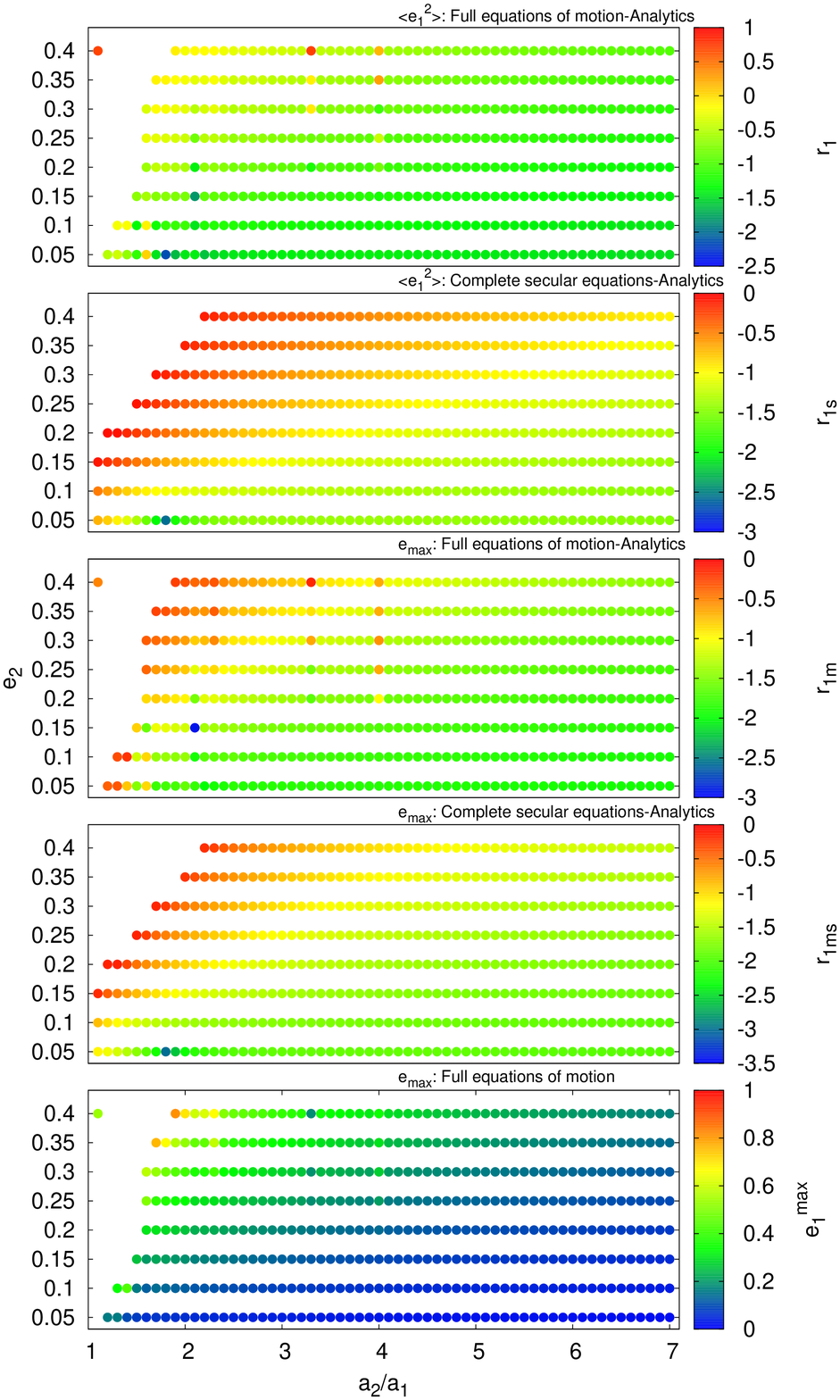}
\includegraphics[width=92mm,height=120mm]{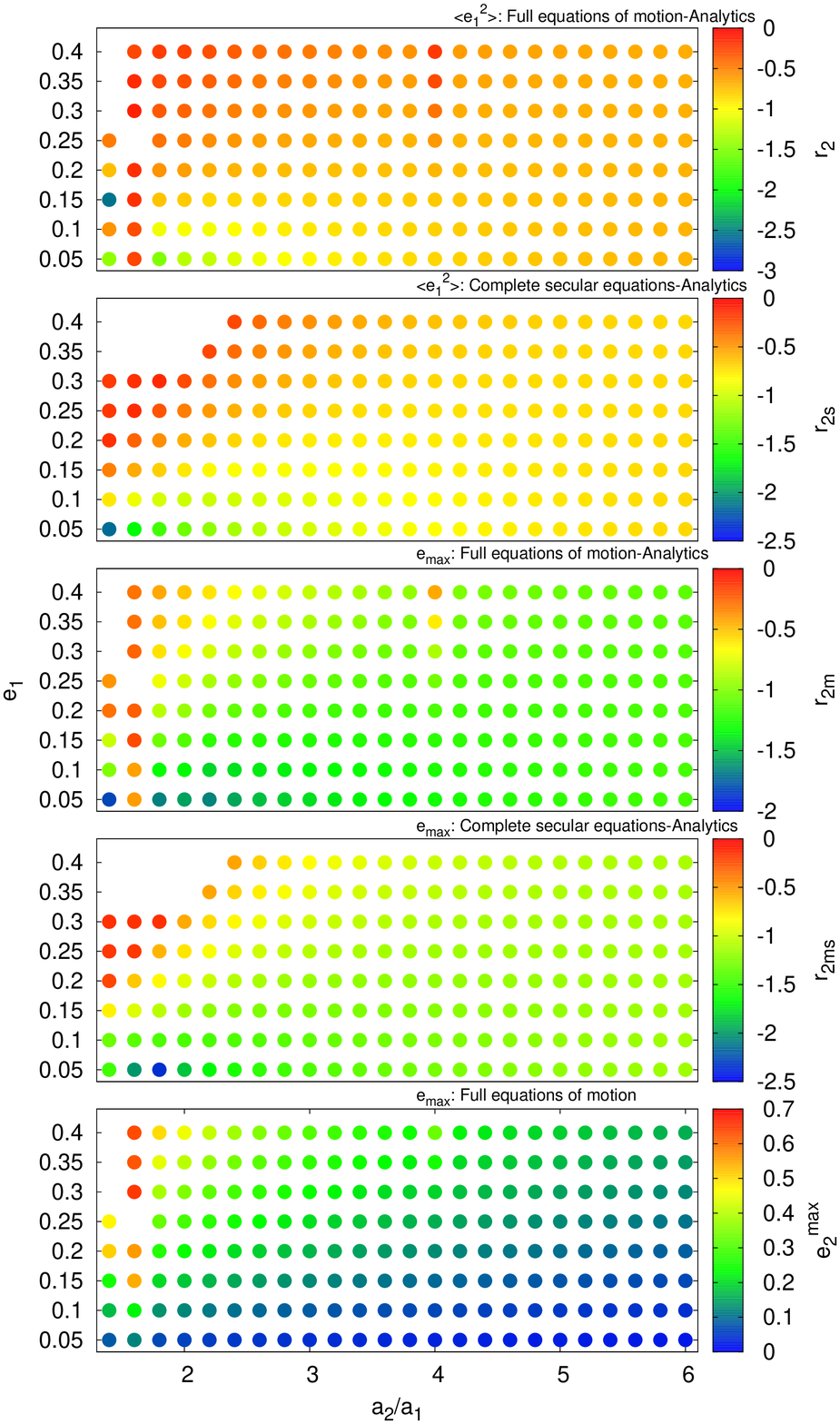}
\caption{Same as in figure 2, but for a system with $m_2=0.1 M_J$ for the inner case (left plot)
and $m_1=0.1 M_J$ for the outer case (right plot).}
\end{figure}
\end{center}

\begin{center}
\begin{figure}
\includegraphics[width=92mm,height=120mm]{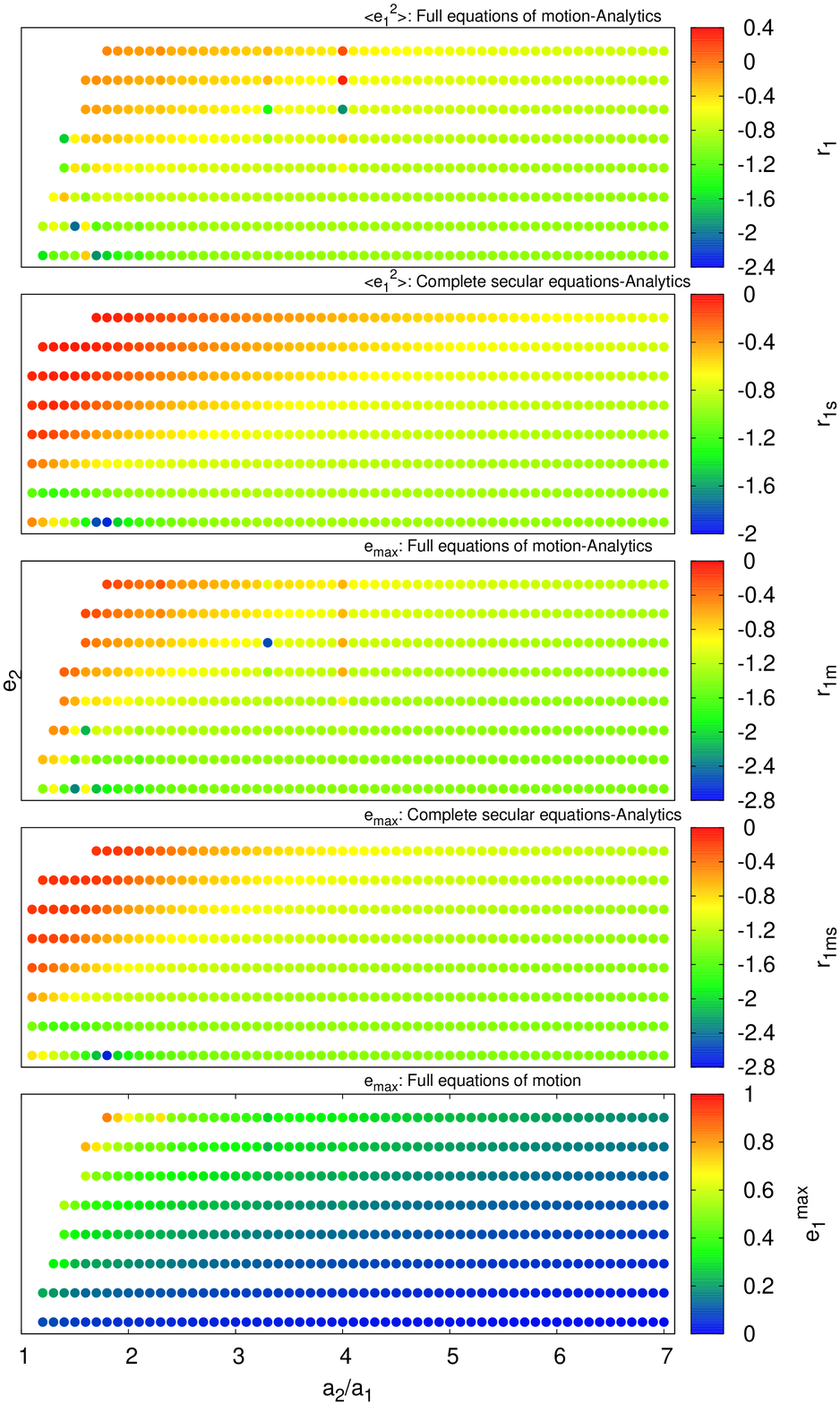}
\includegraphics[width=92mm,height=120mm]{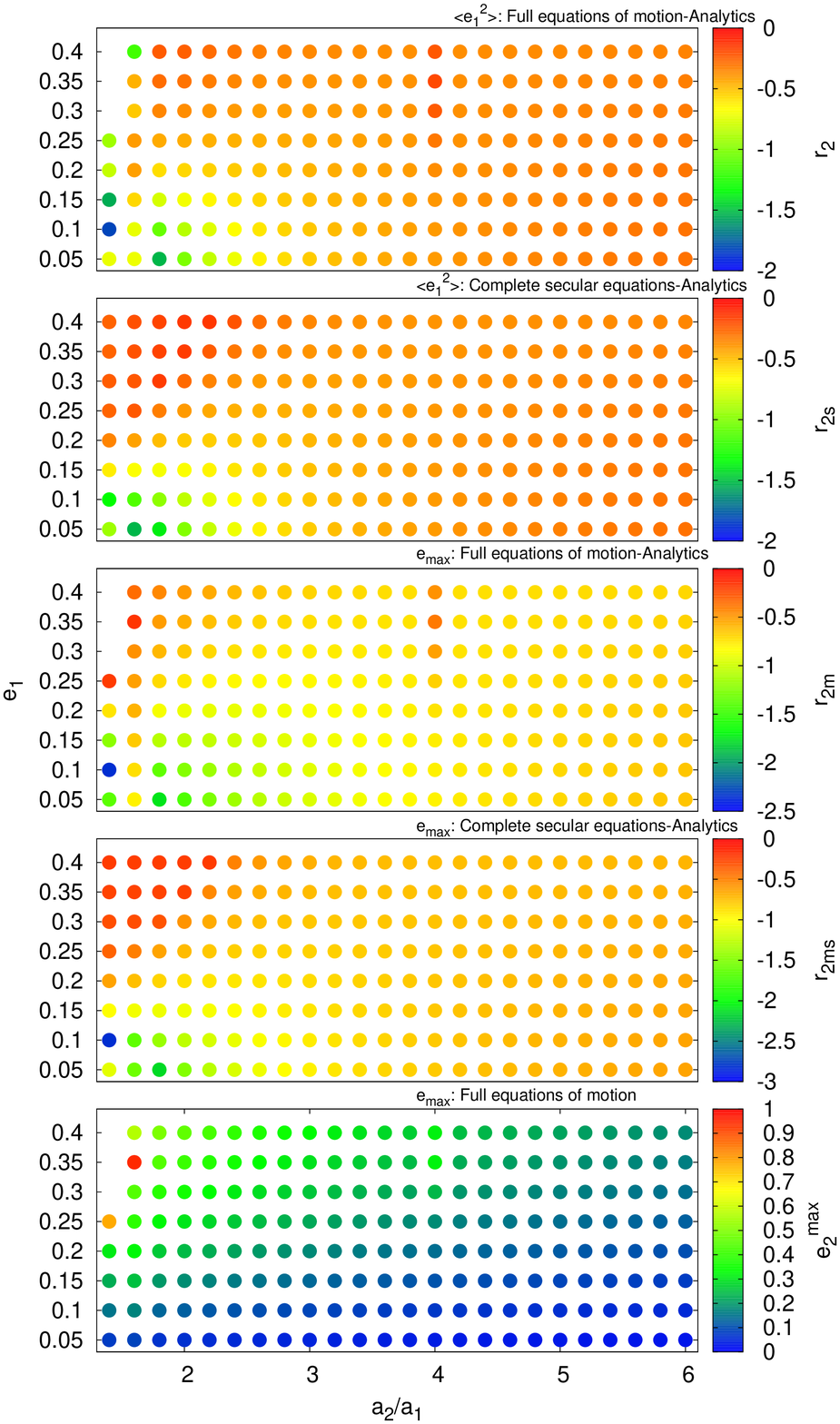}
\caption{Same as in figure 2, but for a system with $m_2=0.03 M_J$ for the inner case (left column)
and $m_1=0.03 M_J$ for the outer case (right column).}
\end{figure}
\end{center}

\bsp	
\label{lastpage}
\end{document}